\documentclass{aastex631}
\usepackage{amsmath}

\submitjournal{Planetary Science Journal}
\shorttitle{Libration-induced Orbit Period Variations}
\shortauthors{Meyer et al.}

\begin{document}

\title{Libration-induced Orbit Period Variations Following the DART Impact}

\correspondingauthor{Alex J Meyer}
\email{alex.meyer@colorado.edu}

\author{Alex J. Meyer}
\affiliation{Smead Department of Aerospace Engineering Sciences, University of Colorado Boulder, Boulder, CO, USA}

\author{Ioannis Gkolias}
\affiliation{Department of Physics, Aristotle University of Thessaloniki, GR 54124 Thessaloniki, Greece}

\author{Michalis Gaitanas}
\affiliation{Department of Physics, Aristotle University of Thessaloniki, GR 54124 Thessaloniki, Greece}

\author{Harrison F. Agrusa}
\affiliation{Department of Astronomy, University of Maryland, College Park, MD, USA}

\author{Daniel J. Scheeres}
\affiliation{Smead Department of Aerospace Engineering Sciences, University of Colorado Boulder, Boulder, CO, USA}

\author{Kleomenis Tsiganis}
\affiliation{Department of Physics, Aristotle University of Thessaloniki, GR 54124 Thessaloniki, Greece}

\author{Petr Pravec}
\affiliation{Astronomical Institute of the Czech Academy of Sciences, Fričova 1, CZ-25165 Ondřejov, Czech Republic}

\author{Lance A. M. Benner}
\affiliation{Jet Propulsion Laboratory, California Institute of Technology, Pasadena, CA 91109-8099, USA}

\author{Fabio Ferrari}
\affiliation{Space Research and Planetary Sciences, Universität Bern, Gesellschaftsstrasse 6, 3012 Bern, Switzerland}

\author{Patrick Michel}
\affiliation{Université Côte d'Azur, Observatoire de la Côte d'Azur, CNRS, Laboratoire Lagrange, CS34229, 06304 Nice Cedex 4, France}

\begin{abstract}

The Double Asteroid Redirection Test (DART) mission will be the first test of a kinetic impactor as a means of planetary defense. In late 2022, DART will collide with Dimorphos, the secondary in the Didymos binary asteroid system. The impact will cause a momentum transfer from the spacecraft to the binary asteroid, changing the orbit period of Dimorphos and forcing it to librate in its orbit. Owing to the coupled dynamics in binary asteroid systems, the orbit and libration state of Dimorphos are intertwined. Thus, as the secondary librates, it also experiences fluctuations in its orbit period. These variations in the orbit period are dependent on the magnitude of the impact perturbation, as well as the system's state at impact and the moments of inertia of the secondary. In general, any binary asteroid system whose secondary is librating will have a non-constant orbit period on account of the secondary's fluctuating spin rate. The orbit period variations are typically driven by two modes: a long-period and short-period, each with significant amplitudes on the order of tens of seconds to several minutes. The fluctuating orbit period offers both a challenge and an opportunity in the context of the DART mission. Orbit period oscillations will make determining the post-impact orbit period more difficult, but can also provide information about the system's libration state and the DART impact.

\end{abstract}


\section{Introduction} \label{sec:intro}
The Double Asteroid Redirection Test (DART) is NASA's first planetary defense mission, and part of the Asteroid Impact and Deflection Assessment (AIDA) collaboration \citep{cheng2018aida}. AIDA is an international collaboration between NASA and ESA, with the purpose of investigating the viability of a kinetic impactor as a means of deflecting a potentially hazardous asteroid \citep{cheng2015asteroid}. The target is the Didymos binary asteroid system, consisting of the primary asteroid Didymos and its smaller secondary moon, Dimorphos. DART will impact Dimorphos as it orbits Didymos, thereby changing the mutual orbital period within the binary asteroid system. The change in orbit period will be measured from ground-based observations in order to calculate the total momentum transferred to the Didymos system. The ground-based observations will be combined with data from LICIACube, a CubeSat carried by DART that will jettison just before impact and observe the collision up-close \citep{dotto2021liciacube}. DART will be followed by Hera, ESA's contribution to AIDA launching in 2024. Hera will rendezvous with the Didymos system in 2026 to complete a post-impact survey \citep{tsiganis2019hera, michel2021esa}. Additionally, Hera will carry two CubeSats, Juventas and Milani, that will aid in the study of the Didymos system, including the probing for the first time of the internal and subsurface properties, and ultimately attempt landings on Dimorphos \citep{goldberg2019juventas, ferrari2021preliminary}.

Binary asteroids are common among near-earth asteroids, comprising at least 15\% of these bodies \citep{pravec2006photometric,pravec1999many}. Binary asteroids are formed by spin-up and fission caused by solar torque of the primary \citep{walsh2008rotational,jacobson2013formation}, and their evolution is then dominated by tidal forces \citep{taylor2010tidal,taylor2014tidal} and the BYORP effect \citep{cuk2005effects,cuk2010orbital,steinberg2011binary}. Tidal forces cause the binary system to expand, while BYORP can either expand or contract the system. Stable binary systems arise when these mechanisms are balanced \citep{jacobson2011long}. In these systems, the primary rotates rapidly and the secondary is tidally locked in a 1:1 spin-orbit resonance. In this work, we assume Didymos in in such an equilibrium. This equilibrium can be disrupted by close planetary encounters \citep{meyer2021effect}.

The relevance of binary asteroids has increased recently due to the interest in these systems as mission targets. Beyond DART, Lucy is visiting the Trojan asteroids \citep{levison2017lucy}, including two binary asteroid pairs \citep{levison2021nasa}. Janus is another mission targeting binary asteroids, also visiting two systems \citep{scheeres2020janus}. This recent uptick in interest can be attributed to the unique advantages binary asteroids provide. Binary asteroids offer an ideal testbed for planetary defense missions due to their fast dynamics \citep{cheng2018aida}, meaning changes in the system are visible from the ground. Binary asteroids also offer new opportunities to study tidal forces and solar radiation torques, since these effects dominate their evolution.

When DART impacts Dimorphos, it will transfer momentum to the Didymos system. This momentum transfer is captured by the momentum enhancement factor $\beta$, which is the ratio of the total momentum transferred to the momentum transferred by a perfectly inelastic collision \citep{cheng2015asteroid}. This captures the effect that ejecta from the surface of Dimorphos will have on the overall momentum transfer. Thus, the minimum value of $\beta$ is 1 corresponding to an inelastic collision with no ejecta and the sole transfer of momentum from the spacecraft to the asteroid. The true value of $\beta$ will be larger than 1 and dependent on the properties of Dimorphos.

The predicted effects of the DART impact have been studied in depth. \cite{agrusa2020benchmarking} showed that the dynamics within the Didymos system are strongly coupled and non-Keplerian, typical of binary asteroids. Additionally, they demonstrated that the DART impact will likely induce significant libration in the secondary. \cite{agrusa2021excited} further examined the induced libration, and pointed out instabilities caused by resonances between the libration frequencies and the mean motion. Additional impact simulations were carried out by \cite{fahnestock2018didymos} to predict the induced libration amplitude post-impact. The Didymos system has also been studied by radar observations to obtain a shape model of the primary, along with estimates of the system mass and secondary shape \citep{naidu2020radar}. Meanwhile, photometric observations have provided constraints on the orbit and spin pole of Didymos \citep{scheirich2019observations, thomas2021constraining}.

Binary asteroids such as Didymos are modeled by the full two body problem (F2BP), in which the rotational and orbital motion of the bodies are coupled together \citep{scheeres2002stability}. The coupling is due to the close proximity of the two asteroids along with their generally asymmetric shapes. The equations of motion for the F2BP were developed by \cite{maciejewski1995reduction} and further studied by \cite{scheeres2006relative}, \cite{tricarico2008figure}, and \cite{scheeres2009stability} among others. Later, \cite{mcmahon2013dynamic} studied the libration in a coupled system which serves as the basis for this analysis. High-fidelity numerical simulations of binary asteroids in the F2BP were made possible by work done by \cite{werner2005mutual}, allowing for the use of polyhedral shape models in the equations of motion. \cite{fahnestock2006simulation} used this method in an in-depth study of 1999 KW4 (now Moshup). Numerical efficiency was improved by the iterative algorithm developed by \cite{hou2017mutual} and implemented by \cite{davis2020doubly} in the General Use Binary Asteroid Simulator (\textsc{gubas}) \citep{davis2021GUBAS}. \textsc{gubas} is used extensively for the DART mission to model the dynamics of Didymos \citep{agrusa2020benchmarking}, and we will use it in this work to verify our results. \textsc{gubas} allows for the gravity field of polyhedron shape models to be truncated at a user-specified degree and order, usually fourth, in order to produce high-fidelity simulations of a binary asteroid.

While previous studies have characterized the coupled dynamics of Didymos \citep{agrusa2020benchmarking,naidu2020radar,hirabayashi2019assessing}, they do not provide the complete picture. In particular, while the libration of Dimorphos has been studied extensively \citep{agrusa2021excited}, the libration's effect on the orbit dynamics has not been investigated in the same level of detail. This analysis attempts to fill this gap by studying the libration and orbit dynamics together rather than separately. In order to do this we rely on the work done by \cite{mcmahon2013dynamic}, who lay the groundwork in examining the coupling between the libration and orbit dynamics. We use results from this study, combined with full numeric \textsc{gubas} simulations, to study the link between induced libration and apparent variations in the mutual orbit period. While the classical two-body problem expects a constant orbit period, this is not the case in the F2BP where the rotation and translation of the bodies are coupled together. The libration of the secondary results in variations in the secondary's spin angular momentum as the secondary slows its rotation at the apex of its libration and speeds up as it passes through $0^{\circ}$. Furthermore, larger libration amplitudes result in larger variations in the spin angular momentum. In order to enforce the conservation of the system's angular momentum, the orbit angular momentum must also vary in response to the secondary's changing angular momentum. This variation in orbit angular momentum results in fluctuations in the orbit period.

The nominal pre-impact Didymos system and DART spacecraft parameters have been established by a variety of previous work in the literature and collected in \cite{agrusa2020benchmarking} and \cite{agrusa2021excited}. Relevant parameters for this work are listed in Table \ref{parameters} for convenience. These values will be used to define the DART spacecraft's mass and velocity and the pre-impact orbit of the Didymos system, unless otherwise noted in the analysis.

\begin{table}[ht]
\begin{center}
 \caption{\label{parameters} The parameters for the nominal Didymos system \citep{agrusa2020benchmarking, naidu2020radar, pravec2006photometric, scheirich2009modeling, agrusa2021excited}.}
\begin{tabular}{c|ccc } 

Symbol & Parameter & Value & Source\\
\hline
$a_{orb}$ & Orbit semimajor axis & 1.19 $\pm$ 0.03 km & Measured \\
$e_{orb}$ & Orbit eccentricity & $<$ 0.03 & Measured \\
$i_{orb}$ & Orbit inclination & 0.0 & Assumed\\
$D_p$ & Primary bulk diameter & 780 $\pm$ 30 m & Measured \\
$D_s$ & Secondary bulk diameter & 164 $\pm$ 18 m & Derived \\
$M_{sys}$ & System mass & (5.37 $\pm$ 0.44)$\times10^{11}$ kg & Derived\footnote{\label{note1}Derived using Kepler's laws, so some error is expected.} \\
$\nu$ & Mass fraction & 0.99 & Derived \\
$P_{orb}$ & Mutual orbit period & 11.9217 $\pm$ 0.0002 h & Measured \\
$P_A$ & Primary spin period & 2.2600 $\pm$ 0.0001 h & Measured \\
$P_B$ & Secondary spin period & 11.9217 $\pm$ 0.0002 h & Assumed \\
$\rho$ & System bulk density & 2170 $\pm$ 350 kg m$^{-3}$ & Derived$^{\text{\ref{note1}}}$ \\
$a/b$ & Secondary major axis ratio & 1.3 & Assumed\footnote{\label{note2}Assumed based on typical binary asteroid secondary shapes.} \\
$b/c$ & Secondary minor axis ratio & 1.2 & Assumed$^{\text{\ref{note2}}}$ \\
$m_{DART}$ & Mass of DART & 535 kg & Assumed \\
$v_{DART}$ & Relative velocity of DART & 6.6 km s$^{-1}$ & Assumed \\

 \hline
 \end{tabular}
 \end{center}
\end{table}

This work is organized as follows. First we present a simplified, analytic dynamics model of a binary asteroid system to develop a link between the libration and orbit period in Section \ref{sec:model}. Section \ref{sec:results} then uses this model, along with \textsc{gubas} simulations for verification, to study how the DART impact induces libration and, in turn, variations in the orbit period. In Section \ref{sec:beta} we test over multiple levels of impact excitation and analyze the system's responses, considering uncertainties in the secondary's shape and configuration. We then relax our simplifications to simulate the full three-dimensional impact in Section \ref{sec:high-fidelity}, using asymmetric shape models. Section \ref{sec:observation} discusses the implications of these results for the observation of Didymos, and, by extension, other binary asteroid systems. We then discuss the results and provide brief conclusions in Section \ref{sec:discussion}.

\section{Dynamics Model} \label{sec:model}
We first develop equations of motion for a simplified binary asteroid system following \cite{mcmahon2013dynamic}. In the simplified system, we constrain the motion to be planar, and model the primary as an oblate spheroid and the secondary as a triaxial ellipsoid. We nominally assume the secondary to be tidally locked pre-impact, as a result of enforcing equilibrium in the system. In this configuration, the long axis of the secondary is always pointing at the center of the primary, and its spin period is exactly equal to the orbit period. The impact will cause deviations from this equilibrium allowing for libration, and, consequently, orbit period variations.

\subsection{Equations of Motion}\label{eom} 
\begin{figure}[ht]
   \centering
   \includegraphics[width = 3in]{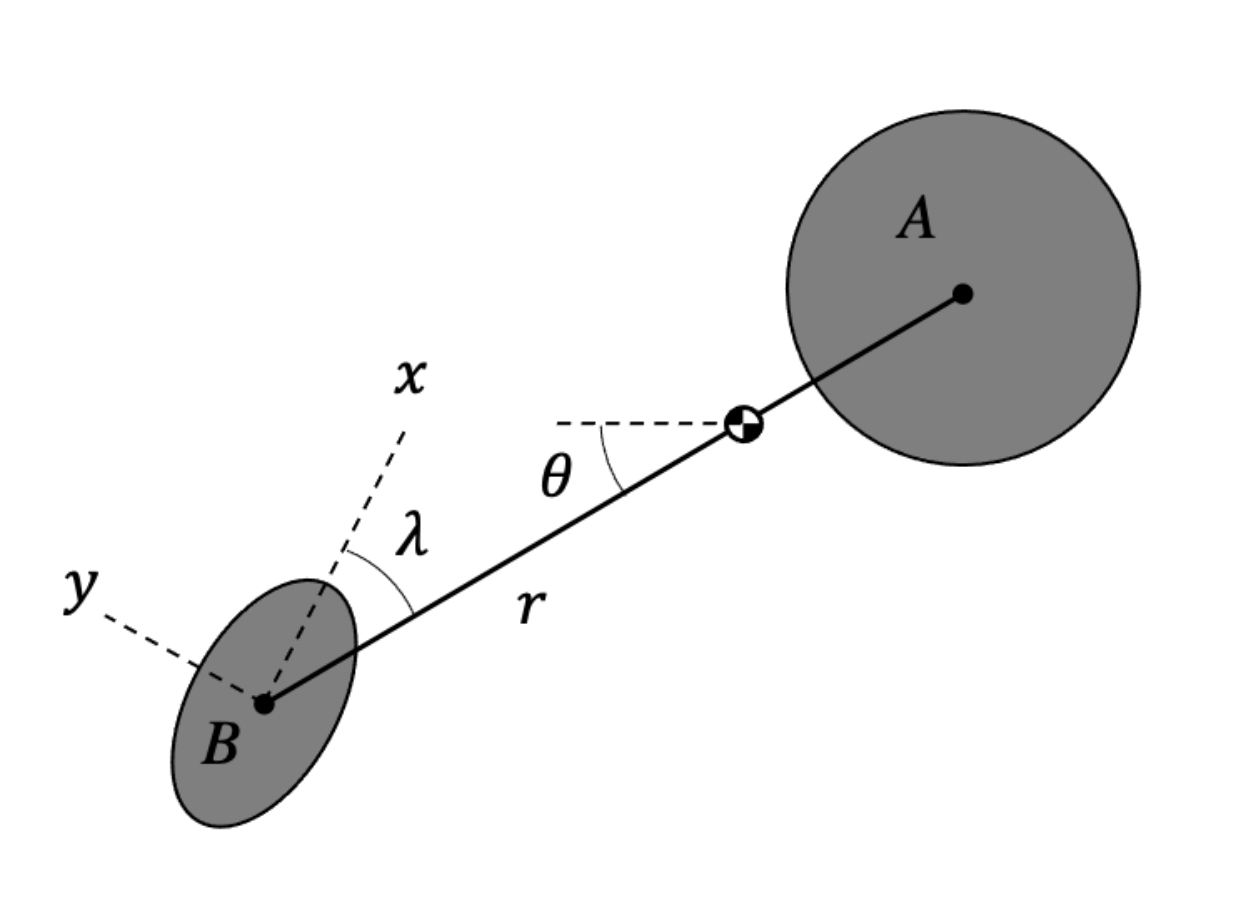} 
   \caption{A top-down view of the simplified, planar model for the binary asteroid. The primary is body $A$ and the secondary is body $B$. In the singly synchronous configuration, $\lambda$ is the libration angle.}
   \label{model}
\end{figure}

In developing equations of motion for the binary system dynamics, we approximate the primary as an oblate spheroid with moments of inertia $I_{A,z}>I_{A,y}=I_{A,x}$ and the secondary as a triaxial ellipsoid with moments of inertia $I_{B,z}>I_{B,y}>I_{B,x}$. Here $A$ refers to the primary and $B$ to the secondary. In each of the body-fixed coordinate frames of the primary and secondary, $x$ is the longest principal body-fixed axis, $y$ is the intermediate principal body-fixed axis, and $z$ is the shortest principal body-fixed axis. Since the primary is an oblate spheroid, for simplicity we define $I_s=I_{A,y}=I_{A,x}$ as the equatorial moment of inertia of the axisymmetric primary. In this dynamics model, we assume all orbital motion lies within the plane and the poles of the primary, secondary, and mutual orbit are all aligned perpendicular to the plane of motion. An illustration of the system is shown in Figure \ref{model}, along with the relevant variables. 

The full system's rotation in inertial space is measured by the angle $\theta$ and the separation of the two bodies' centers of mass is $r$. The orientation of the secondary relative to the rotating line $r$ is measured by $\lambda$, which is also the libration angle. Due to the azimuthal symmetry of $A$ we do not concern ourselves with its rotation.

We normalize the system using the initial separation of the two bodies, denoted $\alpha$, and the mass of the secondary, $m_B$. The time is normalized using the mean motion of the system at the distance $\alpha$: $n=\sqrt{\mathcal{G}(m_A+m_B)\alpha^{-3}}$. The system's mass fraction is defined as $\nu={m_A}(m_A+m_B)^{-1}$, which is always greater than 0.5 and about equal to 0.99 in the nominal system (see Table \ref{parameters}). The normalized inertia tensors are

\begin{equation}
\bar{I}_i = \frac{I_i}{m_i\alpha^2}
\end{equation}
for body $i$, either $A$ or $B$.

In order to develop full equations of motion, we define the potential energy of the system. Here we further simplify the system by only taking the second order expansion of the potential energy. This is the identical expression used in \cite{mcmahon2013dynamic}, so we will skip any derivation and directly report the result:

\begin{equation}
V(r,\lambda) = \frac{-\nu}{r}\bigg[1+\frac{1}{2r^2}\bigg((\bar{I}_{A,z}-\bar{I}_s)-\frac{1}{2}\bar{I}_{B,x}-\frac{1}{2}\bar{I}_{B,y}+\bar{I}_{B,z}+\frac{3}{2}(\bar{I}_{B,y}-\bar{I}_{B,x})\cos2\lambda\bigg)\bigg].
\end{equation}
The potential energy is used in the equations of motion, where we will again skip any derivation:

\begin{equation}
\ddot{r} = \dot{\theta}^2r-\frac{1}{\nu}\frac{\partial V}{\partial r}
\end{equation}

\begin{equation}
\ddot{\lambda} = -\bigg(1+\frac{\nu r^2}{\bar{I}_{B,z}}\bigg)\frac{1}{\nu r^2}\frac{\partial V}{\partial \lambda} + \frac{2\dot{r}\dot{\theta}}{r}.
\end{equation}

In equilibrium, the mutual orbit is circular and the secondary is tidally locked. Thus, the equilibrium conditions are

\begin{equation}
\dot{r}=\ddot{r}=0
\end{equation}
and

\begin{equation}
\lambda=\dot{\lambda}=\ddot{\lambda}=0.
\end{equation}
These conditions allow us to solve for the orbital rate at equilibrium:

\begin{equation}
\dot{\theta} = \sqrt{\frac{1}{\nu r}\frac{\partial V}{\partial r}},
\end{equation}
which is constant. Thus, at equilibrium (i.e. 0$^\circ$ libration), the orbit period is constant.

\subsection{Free Angular Momentum}\label{angmntm} 
Coupling between the libration of the secondary and the orbit period stems from the conservation of angular momentum. To investigate this, we assume the spin angular momentum of the primary will be essentially unchanged by the DART impact. We turn to \textsc{gubas} numerical simulations to verify this assumption. Based on \textsc{gubas} simulations with a $\beta$ factor equal to 3, The secondary's spin angular momentum changes by around 3.6\% while the orbital angular momentum changes by about 1.2\% as a result of the impact. These changes are much larger compared to the primary's spin angular momentum change of around $2\times10^{-4}\%$, and thus we ignore any change in the primary's angular momentum. Despite the primary's large angular momentum, most of the change in momentum does indeed occur in the secondary and orbit. To investigate the momentum exchange, we define the free angular momentum as the total angular momentum minus the contribution by the primary:

\begin{equation}
H_{free} = H_{total} - H_{A}.
\end{equation}
Under our assumption of constant primary angular momentum, $H_{free}$ is constant in the absence of external perturbations. In the simplified system, the free angular momentum can be written as

\begin{equation}
H_{free} = \bar{I}_z\dot{\theta} + \bar{I}_{B,z}\dot{\lambda},
\end{equation}
where the system's mass-normalized polar moment of inertia is defined as

\begin{equation}
\bar{I}_z = \bar{I}_{B,z}+\nu r^2.
\end{equation}
Note the system polar inertia is a function of $r$, and is thus not necessarily constant. We can rearrange this equation to get a direct relationship between the orbital rate and the libration rate:

\begin{equation}
\dot{\theta} = \frac{H_{free} - \bar{I}_{B,z}\dot{\lambda}}{\bar{I}_z}.
\label{eom1}
\end{equation}

Equation \ref{eom1} can be substituted into the equations of motion to eliminate $\dot{\theta}$:

\begin{equation}
\ddot{r} = \frac{(H_{free}-\bar{I}_{B,z}\dot{\lambda})^2r}{\bar{I}_z^2}-\frac{1}{\nu}\frac{\partial V}{\partial r}
\label{eom2}
\end{equation}
\begin{equation}
\ddot{\lambda} = -\bigg(1+\frac{\nu r^2}{\bar{I}_{B,z}}\bigg)\frac{1}{\nu r^2}\frac{\partial V}{\partial \lambda} + \frac{2\dot{r}(H_{free}-\bar{I}_{B,z}\dot{\lambda})}{r\bar{I}_z}.
\label{eom3}
\end{equation}

\cite{mcmahon2013dynamic} also developed an expression for the separation distance required for equilibrium given a value of the free angular momentum. Here we simply invert this equation to solve for the free angular momentum required for a given separation distance to enforce equilibrium:

\begin{equation}
H^* = \bigg[\frac{\nu^2}{r^5}\bigg(r^6+\bigg[\frac{2\bar{I}_{B,z}}{\nu}+\frac{3}{2}\bar{C}_I\bigg]r^4+\bigg[\frac{\bar{I}_{B,z}^2}{\nu^2}+\frac{3\bar{I}_{B,z}}{\nu}\bar{C}_I\bigg]r^2+\frac{3\bar{I}_{B,z}^2}{2\nu^2}\bar{C}_I\bigg)\bigg]^{\frac{1}{2}},
\end{equation}
where

\begin{equation}
\bar{C}_I = -2\bar{I}_{B,x}+\bar{I}_{B,y}+\bar{I}_{B,z}+\bar{I}_{A,z}-\bar{I}_s.
\end{equation}
At this free angular momentum, all equilibrium conditions are met and the system is in a circular, tidally locked orbit with a given separation $r$.

\subsection{Perturbation by the DART Impact}\label{perturbation} 
The DART impact will be retrograde, thus slowing the secondary in its orbit. When the spacecraft impacts Dimorphos it will remove momentum from the system, thus perturbing the free angular momentum away from the equilibrium value. For a planar, head on collision through the body's center of mass, we can define $\beta$ as \citep{agrusa2021excited}:

\begin{equation}
\beta = \frac{-m_B\Delta v}{m_{DART}v_{DART}},
\end{equation}
where $\Delta v$ is the change in the secondary's orbital velocity, $m_{DART}$ and $v_{DART}$ are the mass and relative velocity of the DART spacecraft, respectively. The negative sign is a result of the retrograde impact, causing the secondary to slow in its orbit. Thus, the negative sign serves to keep the change in velocity $\Delta v$ positive for convenience. For this analysis, we assume a DART spacecraft mass of 535 kg and a relative velocity of 6.6 km/s (See Table \ref{parameters}).

Because we are assuming the impact to be head-on through the secondary's center of mass, only the orbital angular momentum is immediately changed by the impulse. Furthermore, we assume the pre-impact system is in an equilibrium configuration, and therefore the mutual orbit is circular. Thus, the impulse to the free angular momentum is equal to

\begin{equation}
\Delta H = \frac{m_Am_B}{m_A+m_B}r\Delta v.
\end{equation}

Substituting our definition for $\beta$ into this expression gives the free angular momentum perturbation as a function of $\beta$:
\begin{equation}
\Delta H = -\beta\nu m_{DART}v_{DART}r,
\label{beta}
\end{equation}

where again the negative sign is due to the retrograde impact. Finally, we add this perturbation to the equilibrium free angular momentum value for use in the equations of motion:

\begin{equation}
H_{free} = H^*+\Delta H.
\end{equation}
Using this expression for the free angular momentum allows $r$, $\dot{\theta}$, and $\lambda$ to oscillate around their equilibrium values, allowing for the exchange of momentum between the orbit and the secondary's spin.

\section{Results} \label{sec:results}
For the following results, we use a nominal configuration for the Didymos system where the initial separation distance $r$ is 1.2 km, the secondary's bulk diameter is equal to 164 m, and the secondary's axis ratios are $a/b=1.3$ and $b/c=1.2$ (See Table \ref{parameters}). Here, $a$ is the largest semi-axis of the secondary, $b$ is the intermediate semi-axis, and $c$ is the smallest semi-axis. The $c$ axis corresponds to the principal moment of inertia and spin axis of the secondary, while the $a$ axis points toward the primary in equilibrium. In normalized units, the equilibrium free angular momentum for the nominal case is about 3.4, resulting in an equilibrium orbit period of about 11.92 hours, consistent with the value in Table \ref{parameters}. Using the equations of motion for the simplified system, we can choose a value of $\beta$ to examine the behavior of the excited state. 

\subsection{Analytic Results}\label{sec:analytic} 
Assuming the system is initially at equilibrium, we have all the initial conditions necessary to integrate the equations of motion. We simulate the impact by instantaneously perturbing the free angular momentum for a given value of $\beta$. When the free angular momentum is perturbed through $\beta$, the system deviates from the equilibrium configuration. The inertia tensor of the secondary is calculated using its bulk diameter and axis ratios, while the inertia tensor of the primary is calculated by approximating the primary's full shape model as on oblate spheroid. Thus, we have all the quantities needed to propagate the equations of motion.

\begin{figure}[ht]
   \centering
   \includegraphics[width = 5in]{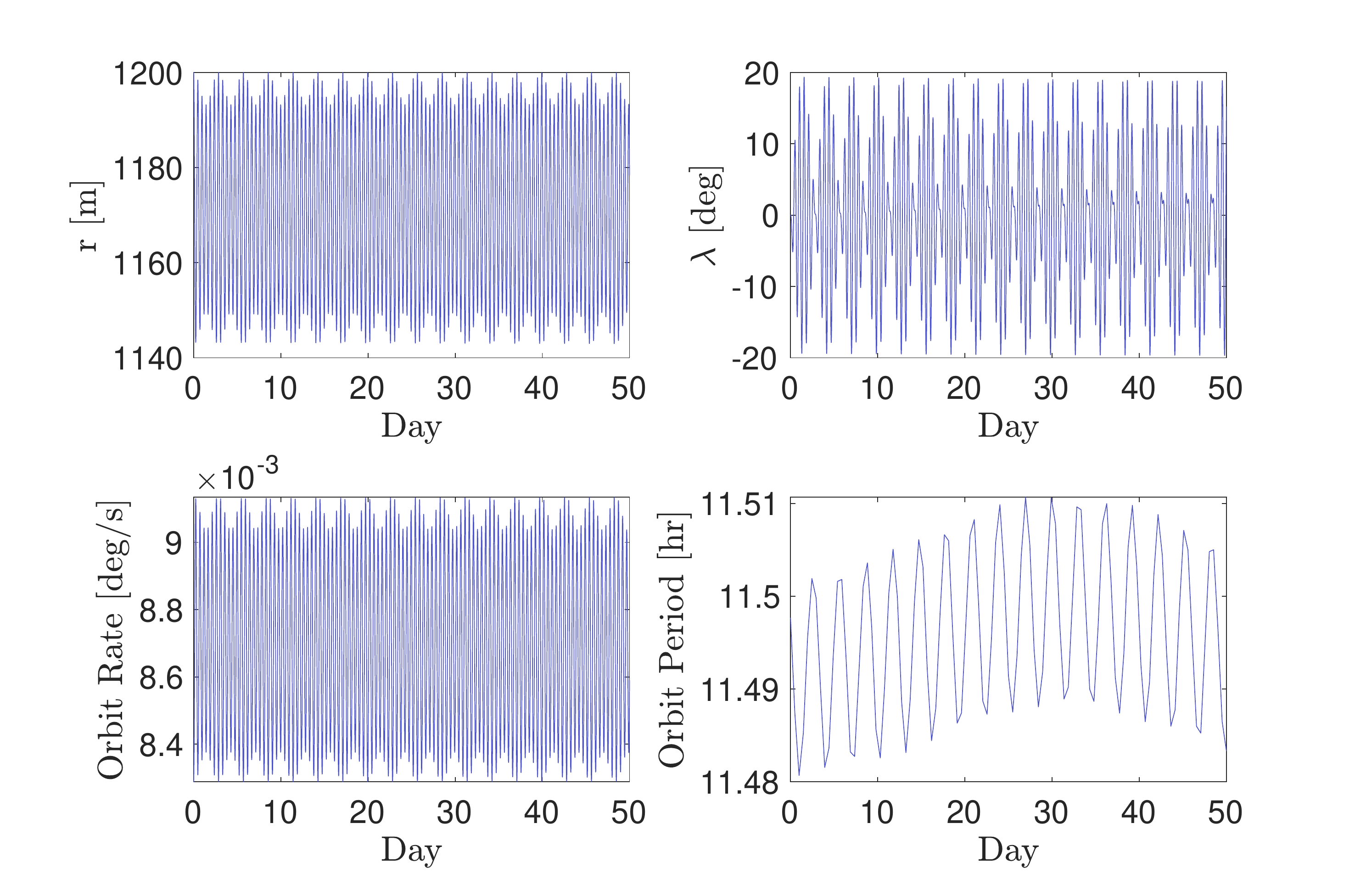} 
   \caption{The separation (top left), libration (top right), orbital rate (bottom left), and orbital period (bottom right) for the case $\beta=3$ for a time of 50 days.}
   \label{analytic_results}
\end{figure}

\begin{figure}[ht]
   \centering
   \includegraphics[width = 5in]{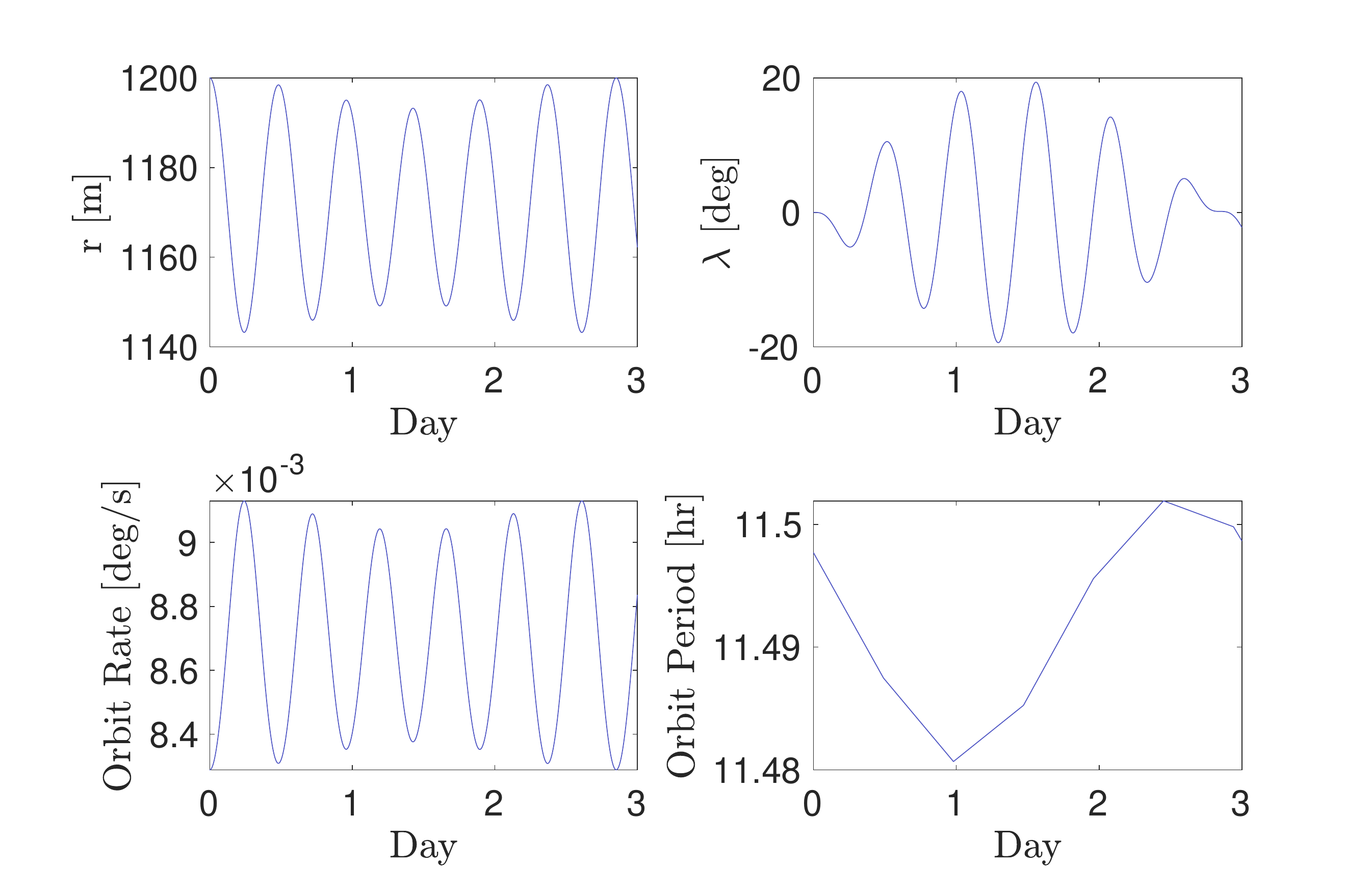} 
   \caption{A closeup view of the separation (top left), libration (top right), orbital rate (bottom left), and orbital period (bottom right) for the case $\beta=3$ for a time of 3 days.}
   \label{analytic_results_zoom}
\end{figure}

We integrate Equations \ref{eom1}, \ref{eom2}, and \ref{eom3} forward in time for a nominal case where $\beta=3$ to solve for $r$, $\lambda$, and $\theta$. Time histories of $r$ and $\lambda$ are shown in the top plots of Figure \ref{analytic_results} for the full 50 day integration period, and a 3-day closeup of these results are shown in Figure \ref{analytic_results_zoom}.

The libration amplitude is calculated using the time history of $\lambda$ over the full time domain. A $20^{\circ}$ libration means the secondary is oscillating by $\pm20^{\circ}$ around the line $r$, as is the case in Figure \ref{analytic_results_zoom}.

The orbit rate $\dot{\theta}$ is plotted for the $\beta=3$ case in the lower left of Figures \ref{analytic_results} and \ref{analytic_results_zoom}, where we immediately see it varies over time. In order to calculate the orbit periods, we use an event function during the integration of the equations of motion. The event function checks the integration of Equation \ref{eom1} and calculates the precise times $\theta$ has reached $2\pi$. By differencing these sequential times, we calculate the orbit period, or the time required for the mutual orbit to move by $2\pi$ radians in inertial space. This is what an ideal lightcurve would see, and therefore is the most accurate representation of real-world dynamics for our model.

 Figures \ref{analytic_results} and \ref{analytic_results_zoom} show the result of this numeric integration approach, plotting the orbit period as a function of time in the lower right plots. These results clearly show that the orbit period is not constant when the system is perturbed from its equilibrium. We calculate the orbit period variation simply by taking the elementary range of the orbit period over its time history, i.e. differencing the maximum and minimum values in the full time domain. Thus, for this nominal case shown in Figure \ref{analytic_results}, the orbit period variation is roughly 0.03 hours, or about 110 seconds. Further discussion on the relevancy of these variations to actual observations is in Section \ref{sec:observation}.

The two timescales between Figures \ref{analytic_results} and \ref{analytic_results_zoom} demonstrate the various modes that drive the dynamics. Estimating these frequencies using a discrete Fourier transform and analyzing the resulting power spectral density, the dynamics of the separation, libration, and orbit rate all have a short-period frequency roughly equal to the system's mean motion, with a period around 12 hours. There is also a long-period mode with a period about equal to 3 days. The frequencies of these modes are dependent on the shape of the secondary and the value of $\beta$ \citep{agrusa2021excited}.

\begin{figure}[ht]
   \centering
   \includegraphics[width = 3in]{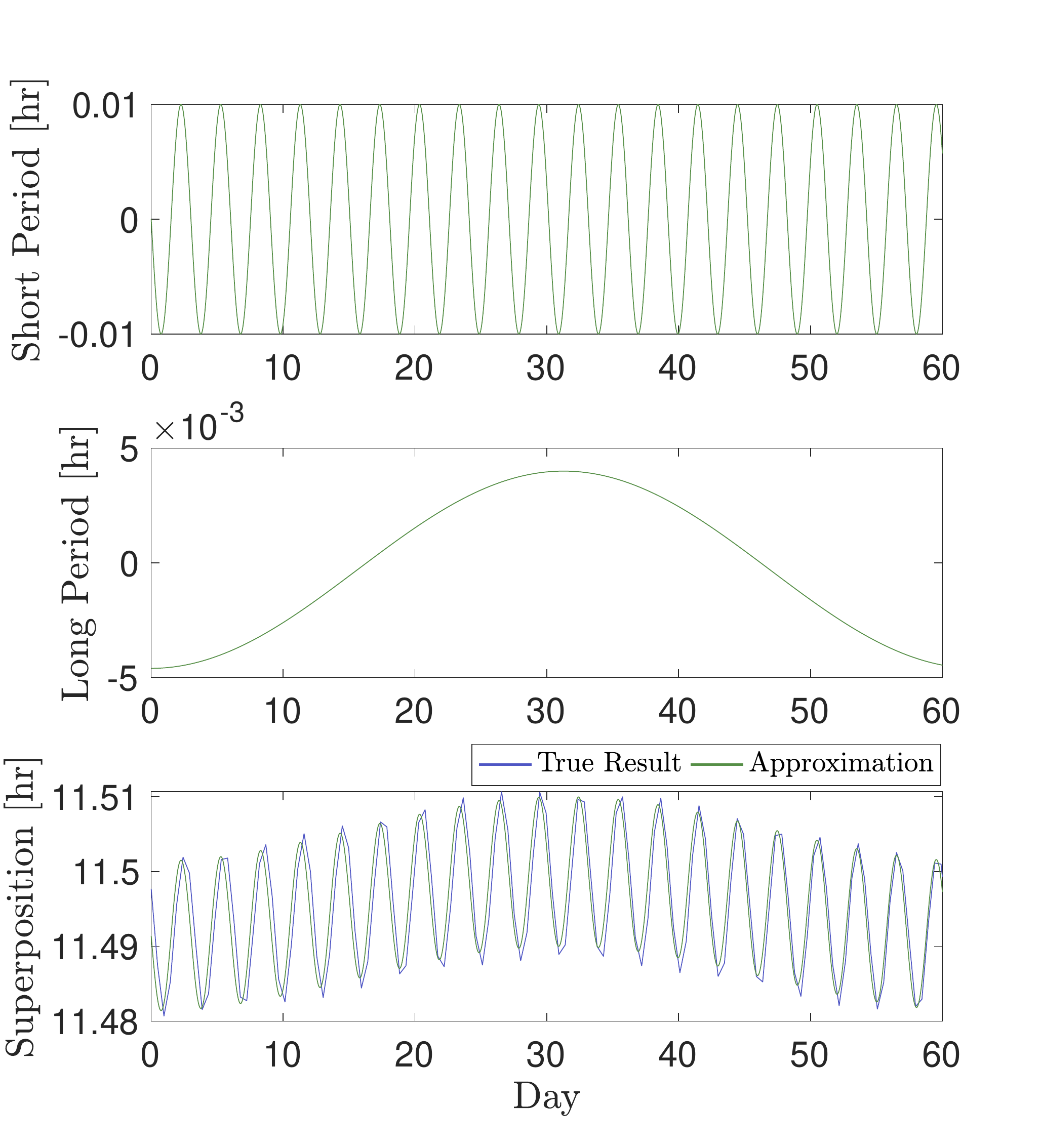} 
   \caption{The decomposition of the orbit period variations into its two constituent modes. The decomposition is computed using a discrete Fourier transform. The short-period mode (top), long-period mode (middle), and the superposition of the two modes (bottom) are plotted, with the superposition overlaid on the actual orbit period calculation.}
   \label{period_decomp}
\end{figure}

Separate from but related to the modes in $r$, $\lambda$, and $\dot{\theta}$, the orbit period fluctuations also appear to be driven by two modes: a short-period and a long-period. Again, we estimate these frequencies by taking a discrete Fourier transform of the time history and analyzing the power spectral density. Here, the short-period mode of the orbit period variations has a period of approximately 3 days, while the long-period mode for the orbit period variations has a period around 60 days. The short-period mode has an amplitude of about 80 seconds, while the long-period mode has an amplitude of about 30 seconds, estimated from the power spectral density. Both of these modes are important, and the total period variation over time is the sum of these two amplitudes. These two modes are plotted in Figure \ref{period_decomp}, and their superposition is also compared to the actual orbit period calculation. Note these modes are only representative, and will change with different secondary shapes and values of $\beta$. A more in-depth discussion on the constituent modes is given in Section \ref{sec:modes}. Since we are mainly interested in the orbit period variations, in the remainder of this work `short-period' and `long-period' will refer to the modes that drive the orbit period, and not those in the separation, libration, and orbit rate.

\subsection{\textsc{gubas} Results}\label{sec:GUBAS}  
In order to propagate the full numeric \textsc{gubas} simulations, we use $\beta$ to calculate the equivalent $\Delta v$ to apply to the system through the relationship in Equation \ref{beta}. The initial conditions are again chosen such that the system is in equilibrium. To achieve this we choose a separation distance of 1200 m and a perpendicular velocity equal to the instantaneous circular orbit speed. However, this does not guarantee the system will have the desired initial orbit period to match Didymos. We therefore iterate the density until the initial orbit period is equal to 11.92 hours, the observed orbit period of Didymos. This is also equal to the spin period of the secondary in the 1:1 spin-orbit resonance. After iteration, the system density is equal to about 2.2 $g/cm^3$, well within the uncertainty on Didymos's density. This results in initial conditions that produce a nearly circular orbit with damped libration and a nearly constant orbit period, while keeping all charateristics of the system within the uncertainties liested in Table \ref{parameters}.

The full numeric simulations then start from the equilibrium initial conditions, with the exception of the velocity perturbation. The \textsc{gubas} results are used for validation of the analytic results, as well as a more accurate representation of the dynamics in Section \ref{sec:high-fidelity}. \textsc{gubas} uses full 3-dimensional simulations with a coupled gravity potential truncated to fourth degree and order along with the full primary shape model to obtain the most accurate model of the Didymos system. We will compare our analytic results to three separate \textsc{gubas} simulations: $\beta=1$, 2, and 3.

In the \textsc{gubas} results, the secondary's orientation is tracked using a set of 1-2-3 Euler angles (roll-pitch-yaw), where $\theta_1$ (roll) measures rotation about the secondary's long axis ($x$-axis), which initially points at the primary. $\theta_2$ (pitch) measures rotation about the secondary's intermediate axis ($y$-axis), and $\theta_3$ (yaw) measures rotation about the secondary's shortest axis ($z$-axis), which is initially aligned with the secondary's spin pole perpendicular to the orbit plane. We are concerned with the planar libration, which is $\theta_3$, or yaw. Rotations in $\theta_1$ and $\theta_2$ are very small in this case, and are consequently ignored. Attitude instabilities are possible depending on the size of the secondary and the value of $\beta$, in which case $\theta_1$ and $\theta_2$ are important \citep{agrusa2021excited}. However, for now we only consider cases where the attitude remains stable. We measure the libration amplitude only using the time history of $\theta_3$.

Calculating the orbit period is slightly different: each time the secondary crosses a pre-defined inertial plane in the \textsc{gubas} results, the time of crossing is recorded. Generally, we simply define this plane to be the inertial $y-z$ plane. Each of these crossing times is then differenced to calculate the time between the plane passages, similar to the time between occultations in actual binary asteroid observations. This is analogous to our approach in the analytic case where we are  concerned with a physical interpretation of the orbit period. We calculate the variation in the orbit period again by differencing the maximum and minimum values of the orbit period over the full time domain.

\section{Sweep Over Momentum Enhancement Factor}\label{sec:beta}
In order to examine the system's response to the DART impact, we will sweep through multiple values of the momentum enhancement factor $\beta$. The minimum value of $\beta$ is 1 and we increase this up to a value of $\beta=5$. In this way we investigate how increasing the momentum transfer of the impact affects the resulting dynamics.

\subsection{Mean Orbit Period Change}\label{sec:periodchange} 
\begin{figure}[ht]
   \centering
   \includegraphics[width = 3in]{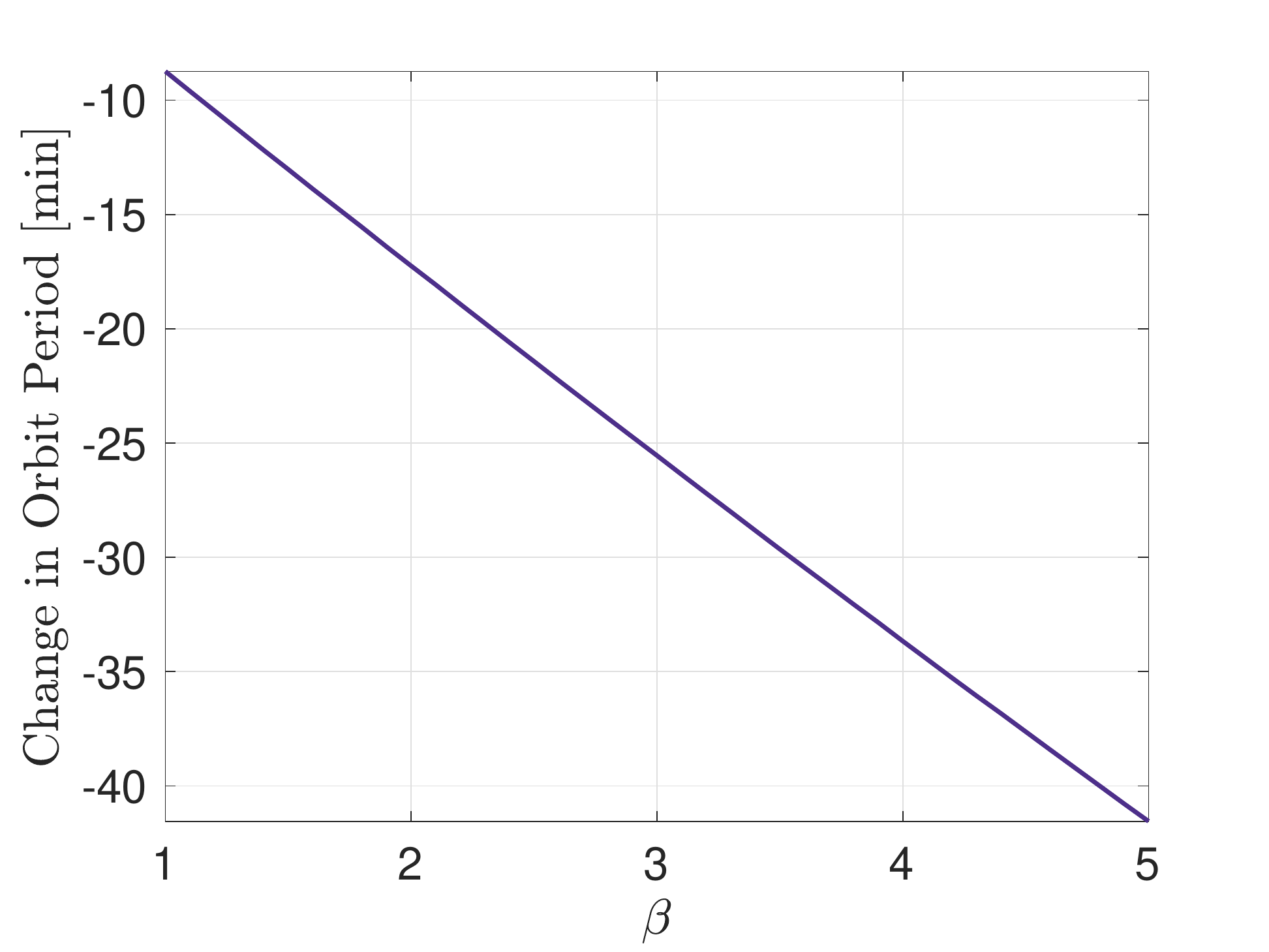} 
   \caption{The change in mean orbit period as a function of the $\beta$ parameter.}
   \label{period_v_beta}
\end{figure}

By calculating the average orbit period following the impact we can obtain a single mean orbit period ignoring the variations. We have already calculated the initial equilibrium orbit period, and by differencing these we can estimate the expected change in mean period for a value of $\beta$. We can test over multiple values of $\beta$, here ranging between 1 and 5, to plot the change in orbit period as a function of $\beta$. This is shown in Figure \ref{period_v_beta}.

This is one of the expected DART results: a change in orbit period caused by the impact. This is the result that will be measurable from ground-based observations. A strongly linear relationship between $\beta$ and the change in orbit period is evident, with a negative slope resulting from the retrograde impact. This shows the minimum value of $\beta=1$ changes the orbit period by approximately 9 minutes, or about 1\% of the equilibrium orbit period.

\subsection{Libration-induced Variations in Orbit Period}\label{sec:periodvariation} 
\begin{figure}[ht]
   \centering
   \includegraphics[width = 3in]{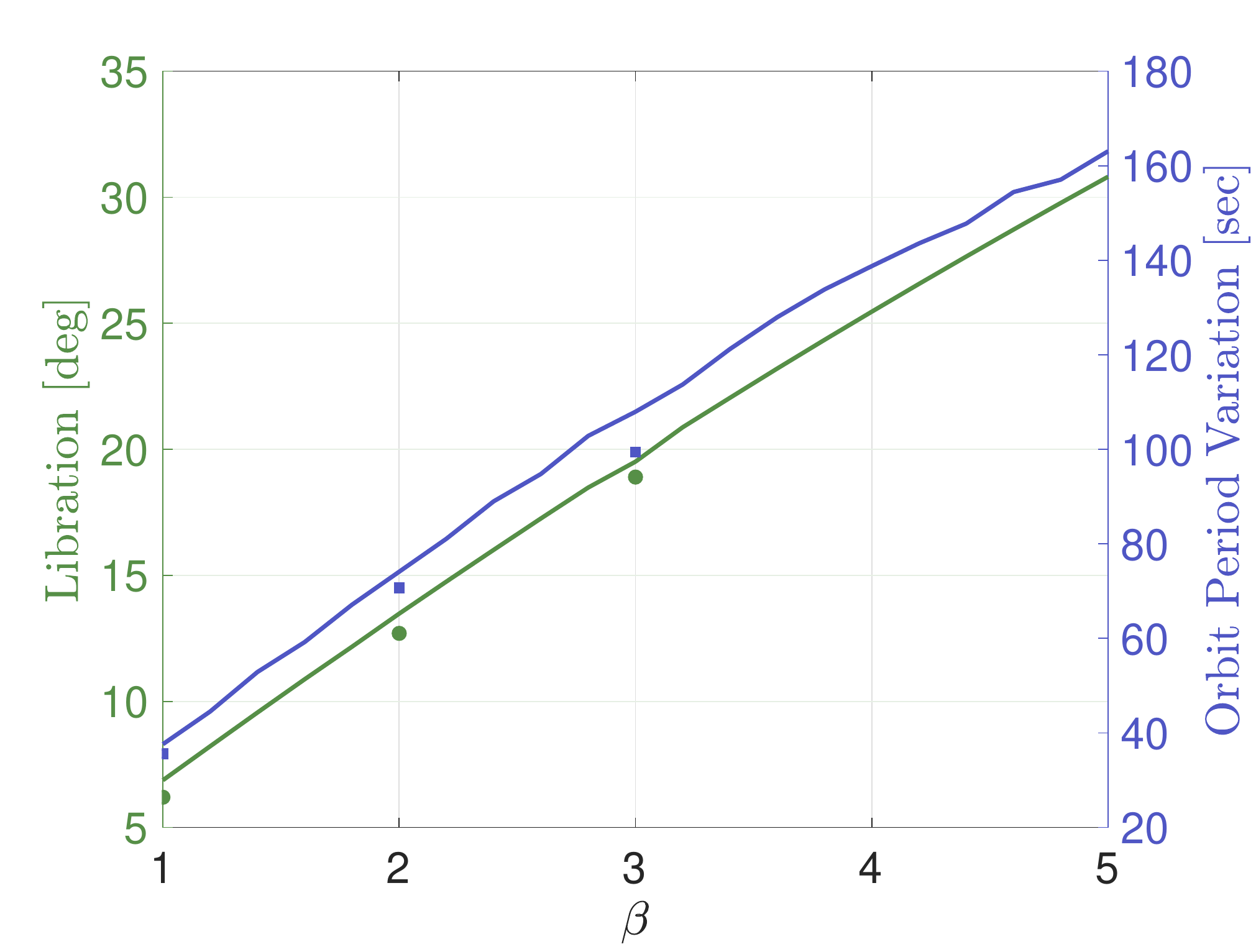} 
   \caption{The libration and orbit period variations as functions of $\beta$. The left axis (green) indicates the libration amplitude and the right axis (blue) shows the variation in the orbit period. The results from fully coupled \textsc{gubas} numeric simulations are overlaid as data points, and agree well with the analytic model.}
   \label{beta_lib_period}
\end{figure}

While DART will clearly change the mean orbit period, it will also cause the orbit period to fluctuate about the mean by inducing librations in the secondary. Using the analytic model, we test multiple values of $\beta$ ranging between 1 and 5 and calculate the resulting libration amplitude and orbit period variations as detailed in Section \ref{sec:analytic}. In addition, several \textsc{gubas} simulations are included for $\beta$ values of 1, 2, and 3 for comparison and validation. The analytic and numeric results are all shown in Figure \ref{beta_lib_period}.

Figure \ref{beta_lib_period} shows that even the smallest value of $\beta=1$ has fluctuations in its orbit period of about 40 seconds. That increases to more than a minute and a half at a $\beta$ value of 3. Thus, even modest libration can result in significant variations in the orbit period. These results agree well with selected \textsc{gubas} simulations, which are shown as data points in Figure \ref{beta_lib_period}

\begin{figure}[ht]
   \centering
   \includegraphics[width = 3in]{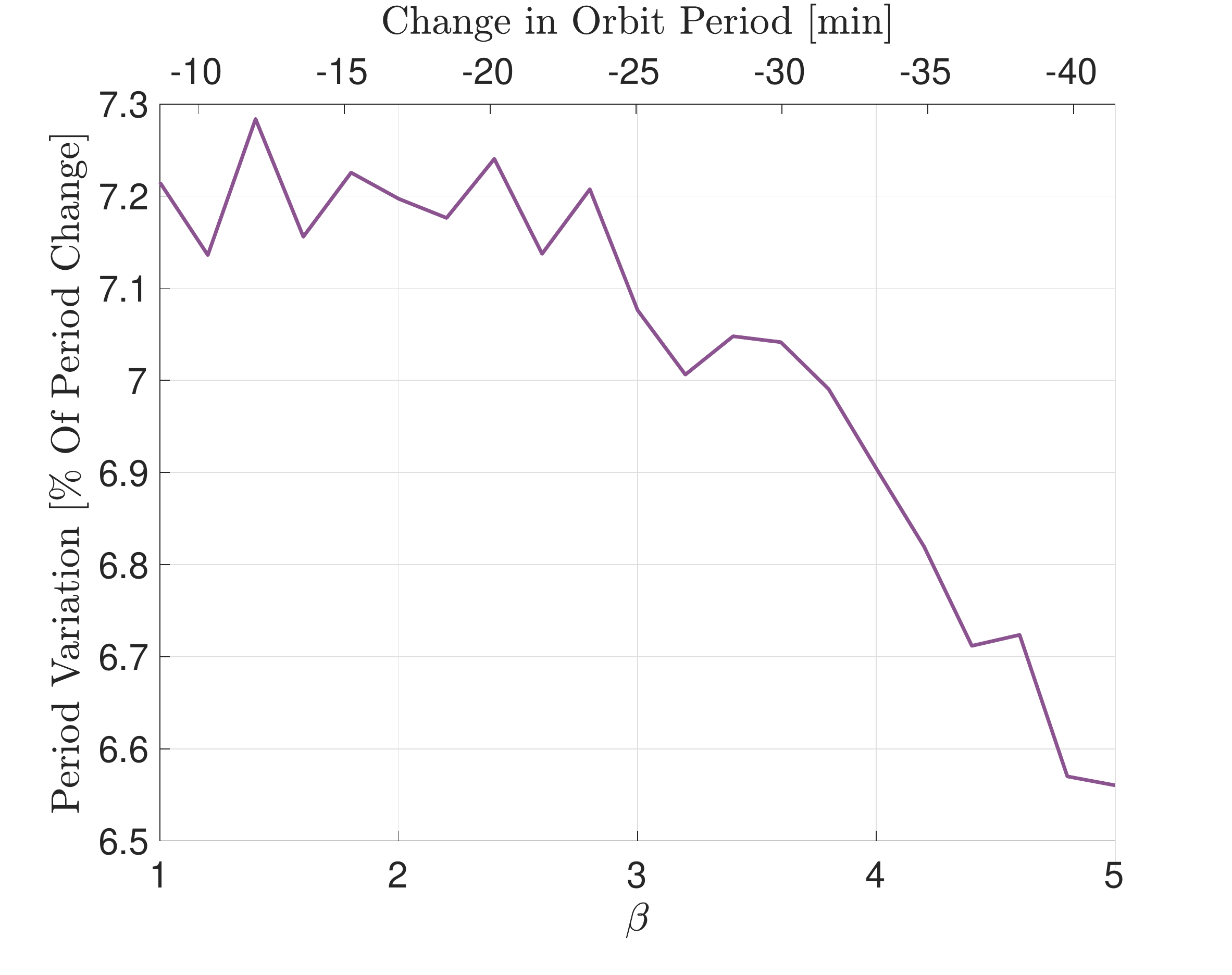} 
   \caption{The variation in orbit period as a percent of the change in orbit period for varying values of $\beta$. The curve is also plotted against the actual change in orbit period.}
   \label{percentperiod}
\end{figure}

Next, we calculate the orbit period variation as a percentage of the change in mean orbit period. This is plotted in Figure \ref{percentperiod}, showing that the smallest values of $\beta$ have the largest relative variations compared to the impulsive change to the mean orbit period. For example, at $\beta=1$, there is only about a 9 minute change to the mean orbit period, but the orbit period fluctuates by around 40 seconds, which is about $7\%$ of 9 minutes. Compared to $\beta=5$, where the mean orbit period changes by more than 40 minutes but sees fluctuations on the order of 160 seconds, which is only about $6.5\%$ of the period change.

\subsection{Effect of Axis Ratio}\label{sec:axisratio} 
\begin{figure}[ht]
   \centering
   \includegraphics[width = 3in]{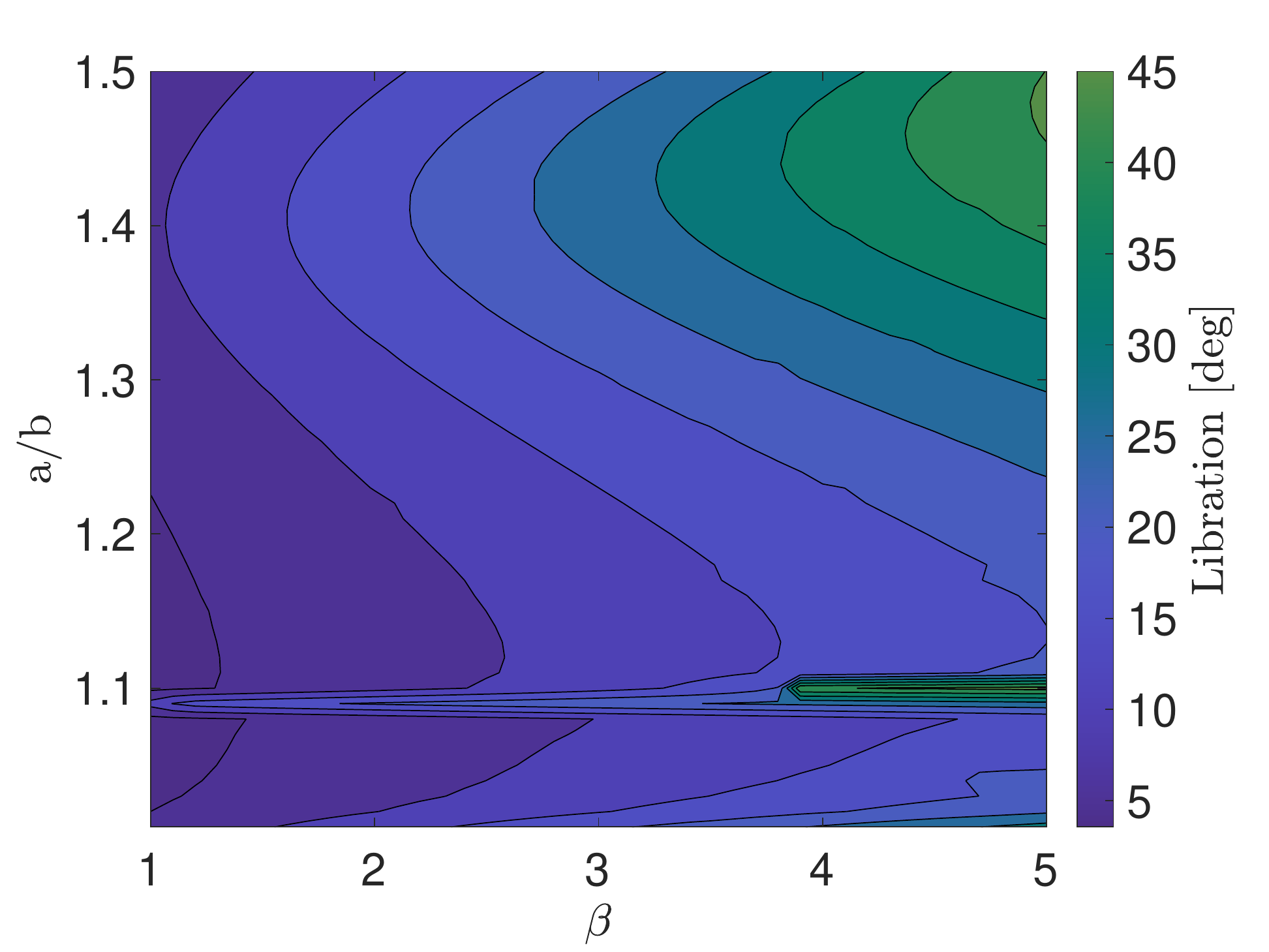} 
   \caption{The libration amplitude as a function of the axis ratio $a/b$ and the $\beta$ parameter.}
   \label{ab_libration}
\end{figure}

\begin{figure}[ht]
   \centering
   \includegraphics[width = 3in]{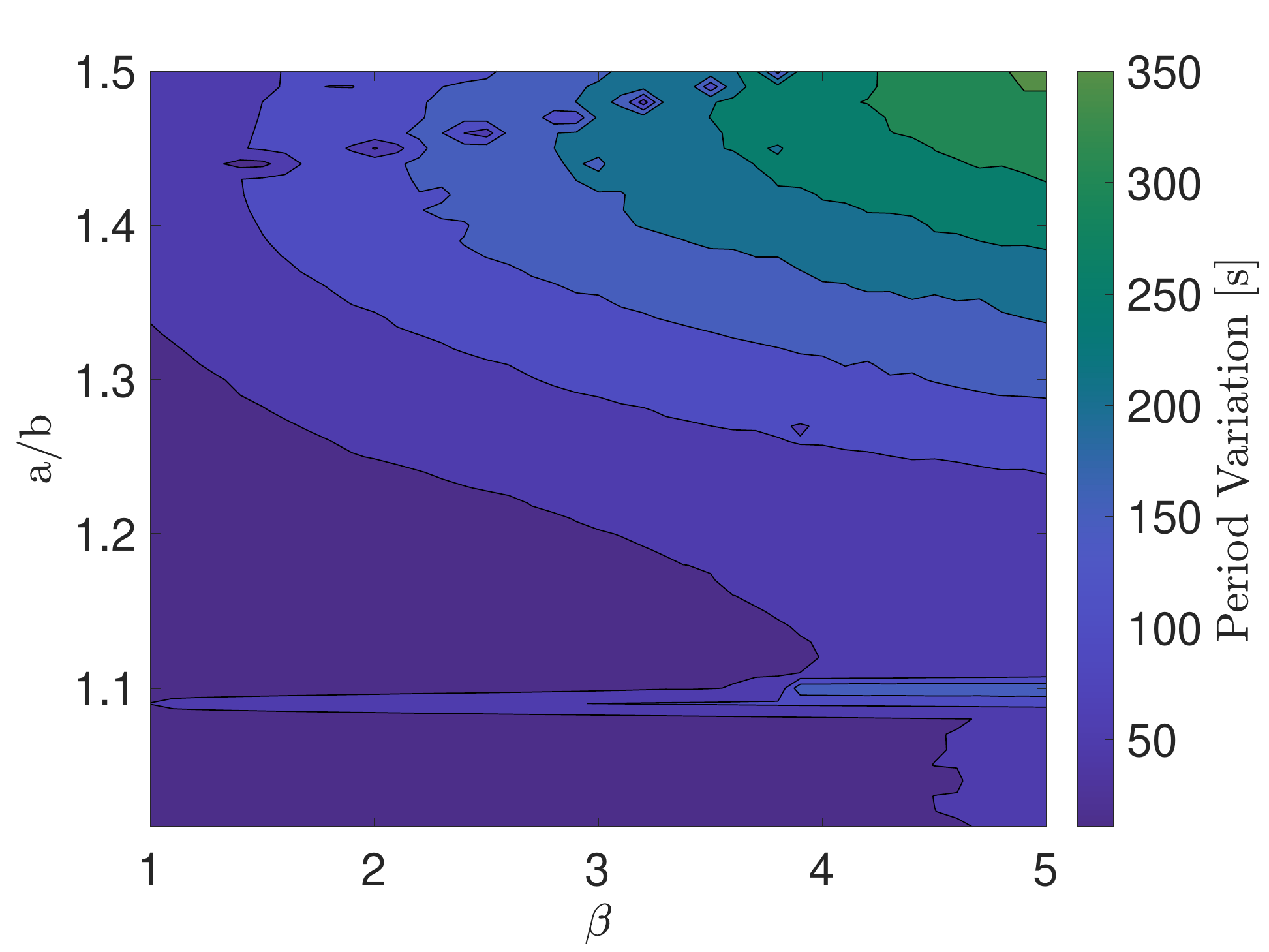} 
   \caption{The variation in orbit period as a function of the axis ratio $a/b$ and the $\beta$ parameter.}
   \label{ab_period}
\end{figure}

While up to now we have only focused on the nominal case $a/b=1.3$ and $b/c=1.2$, it is worthwhile to investigate the effect of the axis ratios on the behavior of the system. While $b/c$ does not have a large effect on the secondary's response due to the planar dynamics, $a/b$ does affect the dynamics significantly. Thus, we will test additional values of $a/b$ while holding $b/c$ constant. We allow $a/b$ to vary between 1.0 and 1.5, the typical range of secondary axis ratios \citep{pravec2016binary}, while still varying $\beta$ between 1 and 5. Figure \ref{ab_libration} shows the libration amplitude as a function of both $a/b$ and $\beta$, while Figure \ref{ab_period} shows the corresponding variation in the orbit period.

Still immediately apparent is that increasing $\beta$ increases both the libration amplitude and the orbit period variations. However, we also notice a strong dependence on $a/b$. For example, there is a resonance near $a/b=1.1$, where the libration and orbit period variation are excited for all values of $\beta$. This is a 2:1 resonance between the libration frequency and mean motion, as discussed in \cite{agrusa2021excited}. The 1:1 resonance between these frequencies occurs near $a/b=1.4$, which is visible in these plots but not as sharp as the resonance at $a/b=1.1$.

As $a/b$ changes, the moments of inertia of the secondary change as well. Also, the natural frequencies of oscillations in the secondary change with $a/b$ \citep{agrusa2021excited}. Thus, it is unsurprising that the libration magnitude changes with $a/b$. However, it is interesting that there is not a monotonic dependence on $a/b$. As an explanation, note that certain shapes excite resonances with the mean motion. While the 2:1 resonance at $a/b=1.1$ appears to be very narrow, the width of the 1:1 resonance near $a/b=1.4$ is much wider. This is consistent with the findings in \cite{agrusa2021excited}, and provides one explanation for the complicated relationship between the libration and $a/b$.

\begin{figure}[ht]
   \centering
   \includegraphics[width = 3in]{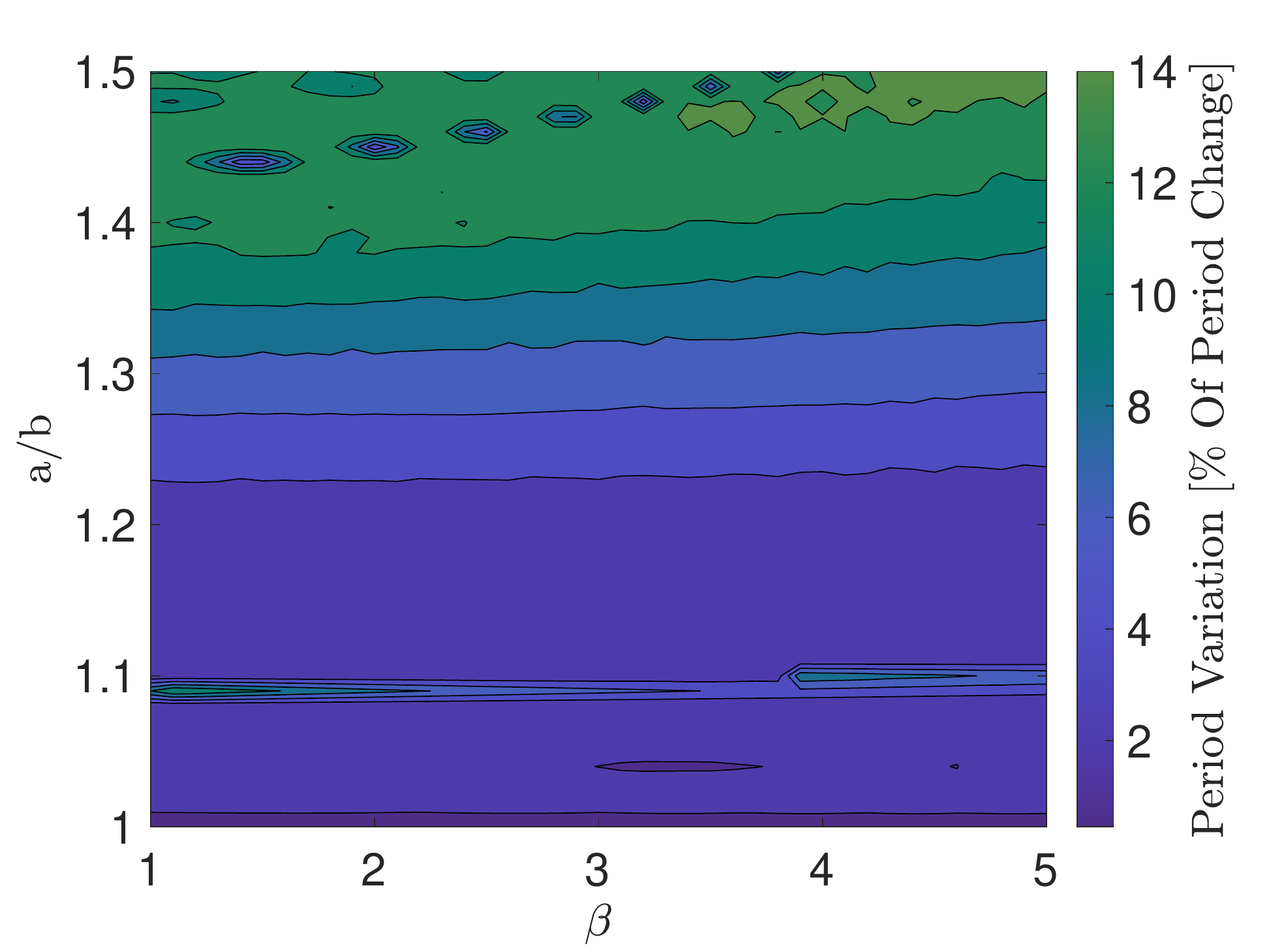} 
   \caption{The variation in orbit period as a percentage of the mean period change for values of the axis ratio $a/b$ and the $\beta$ parameter.}
   \label{ab_percentperiod}
\end{figure}

Lastly, Figure \ref{ab_percentperiod} shows the variations in the orbit period as a percent of the mean orbit period change. This is the same approach taken in Figure \ref{percentperiod}, so that Figure \ref{percentperiod} is a single line in Figure \ref{ab_percentperiod}. The mean period change does not depend on the axis ratio $a/b$. However, increasing $a/b$ does increase the variation in the orbital period.

\subsection{Non-equilibrium Impact}\label{sec:noneq} 
\begin{figure}[ht]
   \centering
   \includegraphics[width = 3in]{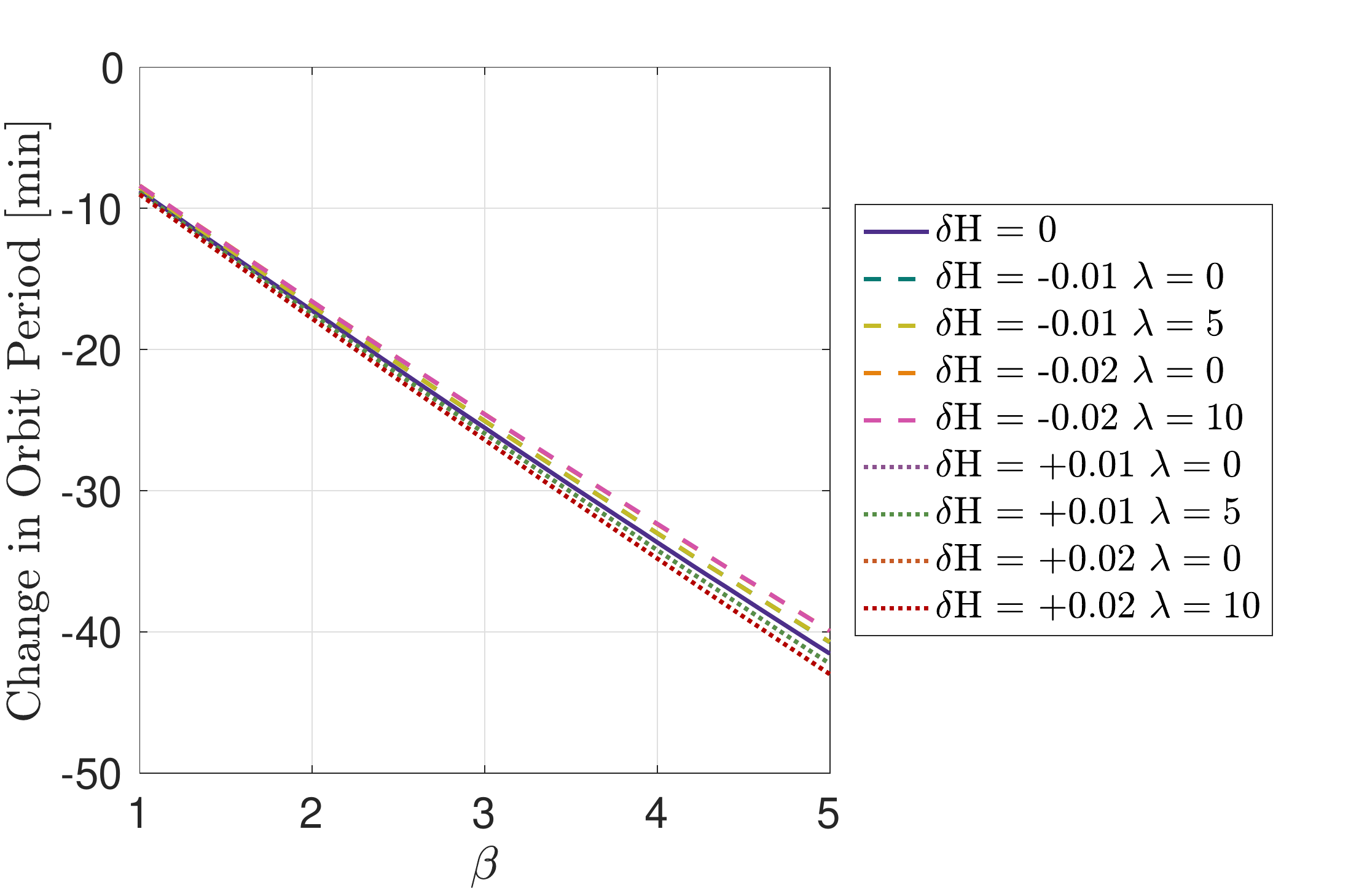} 
   \caption{The overall average period change as a function of $\beta$ for the initial equilibrium and several perturbed cases.}
   \label{noneq_change}
\end{figure}

Up to this point we have assumed Didymos to be in an equilibrium configuration prior to impact. In this state the secondary is perfectly tidally locked and the libration before impact has been damped to $0^{\circ}$. However, it is possible Didymos is not exactly in an equilibrium state before the DART impact. We now examine the post-impact behavior for a system that was not previously in an equilibrium. We test four different excitation levels, perturbing the equilibrium free angular momentum by $\pm$0.01 and $\pm$0.02 in normalized units (corresponding to maximum libration amplitudes of $5^{\circ}$ and $10^{\circ}$, respectively). Two of these cases have a free angular momentum higher than the equilibrium value and the other two lower than the equilibrium value. At each of these levels of perturbation, we run two cases: one with the libration amplitude at the impact epoch equal to $0^{\circ}$, and the other with the libration amplitude at its maximum value for the excitation level, either $5^{\circ}$ or $10^{\circ}$. Thus, we have 8 perturbed test cases in addition to the equilibrium case. First, the overall average period change is shown in Figure \ref{noneq_change}. This demonstrates that the average period change is largely unaffected by impacting a non-equilibrated system. However, systems with a free angular momentum higher than the equilibrium see slightly larger changes in their orbit period, whereas systems below the equilibrium free angular momentum see slightly smaller changes. The effect is more pronounced further away from the equilibrium. The orientation of the secondary at impact has no bearing on the mean period change, regardless of the level of system excitation.

\begin{figure}[ht]
   \centering
   \includegraphics[width = 3in]{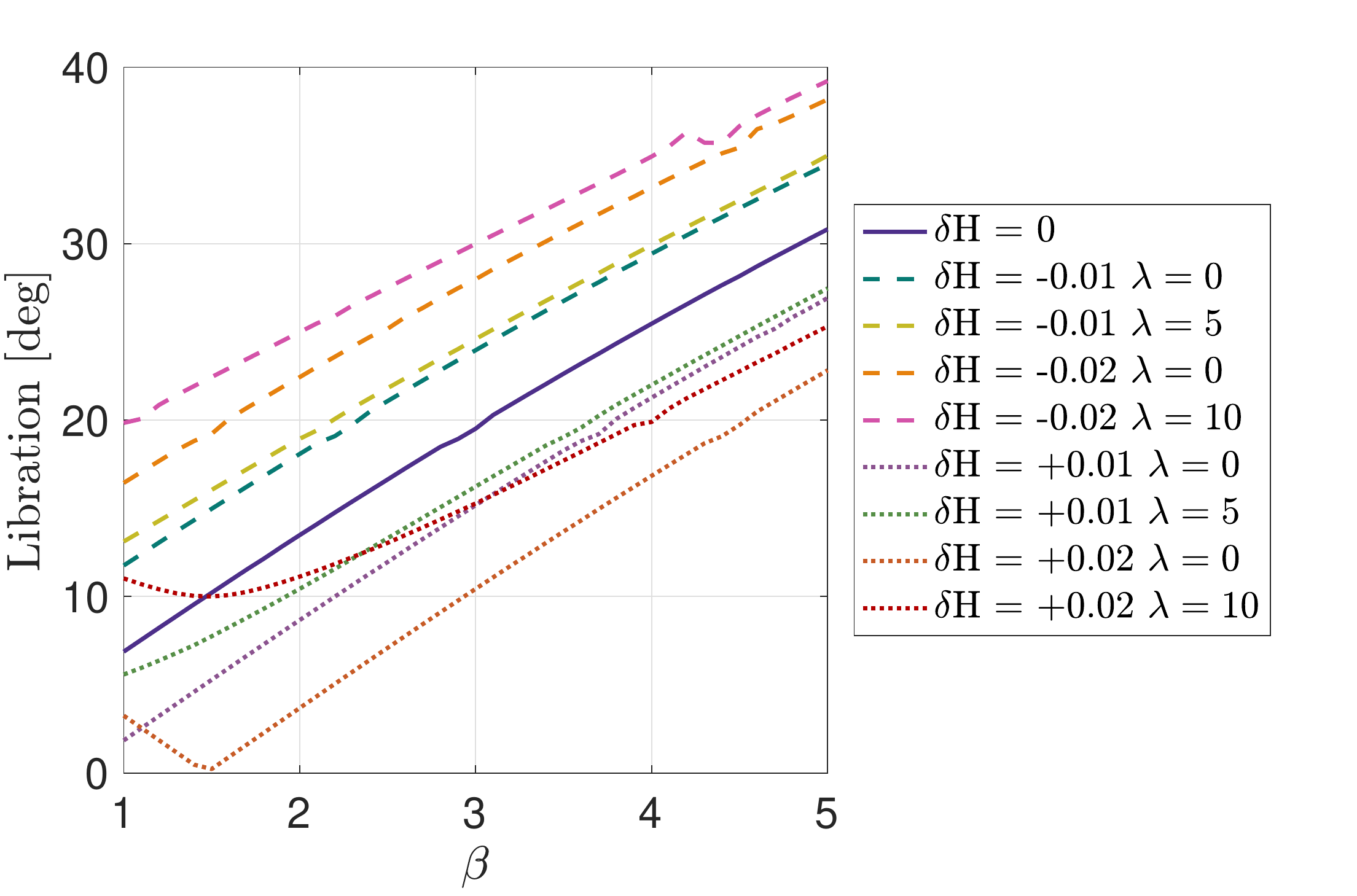} 
   \caption{The libration amplitude as a function of $\beta$ for the initial equilibrium and several perturbed cases.}
   \label{noneq_libration}
\end{figure}

Figure \ref{noneq_libration} shows the libration amplitude as a function of $\beta$ for all test cases. We see clear differences between the cases above the equilibrium and cases below the equilibrium. In general, systems with an initial configuration below the equilibrium angular momentum have larger resulting libration amplitudes following the impact. This is expected, since the impact will move the system further from equilibrium. Additionally, impacts when the secondary is at its apex of libration see a more energetic response than impacts at 0$^\circ$ libration. Systems with an initial angular momentum larger than the equilibrium value see a more diverse range of responses. For systems only slightly perturbed above the equilibrium, even the smallest impact is enough to decrease the angular momentum back through the equilibrium value to a value below it. For systems that are initially more excited, we see that increasing the impact momentum transfer pushes the system back toward equilibrium and the libration amplitude begins to decrease as $\beta$ increases. However, once $\beta$ is large enough to push the system back below equilibrium, the libration amplitude then begins to increase as $\beta$ increases. Again we see a more energetic response when the impact occurs at the secondary's libration apex than when the impact occurs at 0$^\circ$ libration.

\begin{figure}[ht]
   \centering
   \includegraphics[width = 3in]{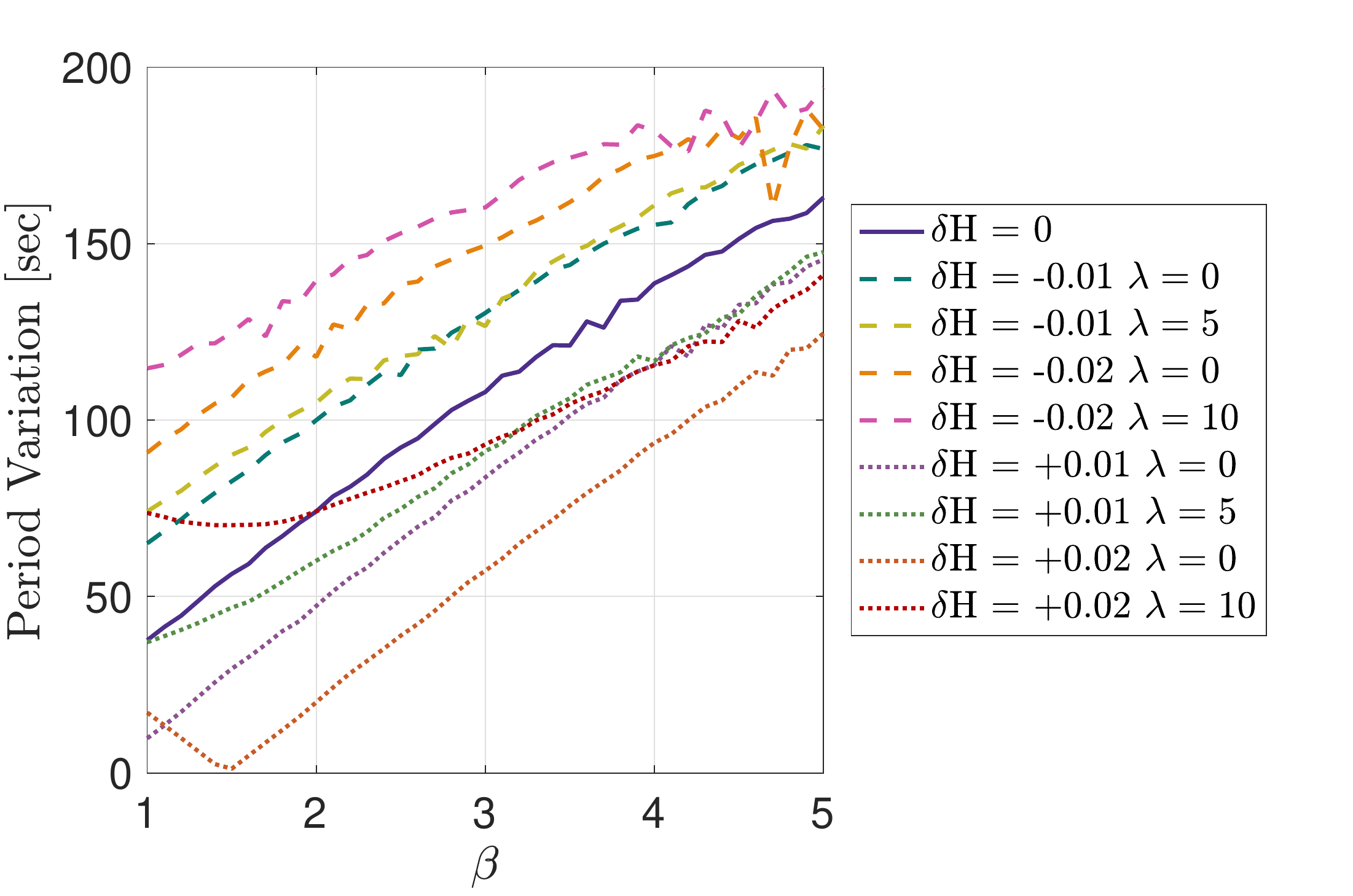} 
   \caption{The variation in orbit period as a function of $\beta$ for the initial equilibrium and several perturbed cases.}
   \label{noneq_period}
\end{figure}

Figure \ref{noneq_period} shows the variation in orbit period for the same test cases. Here we see very similar behavior to the libration amplitude response. Systems initially below the equilibrium angular momentum are pushed further away from equilibrium, whereas systems above the angular momentum are first pushed back toward equilibrium, then further below it as $\beta$ increases. This further reinforces the idea that the libration amplitude and orbit period variations have nearly the same trend as a function of $\beta$.

Another concern for an initially non-equilibrium system is out-of-plane motion, where the secondary rocks about its long axis, as this can be long-lived in binary asteroids \citep{quillen2021non}. This out-of-plane rotation would also cause the system to be initially perturbed from the equilibrium value of angular momentum. However, the perturbation is small since the out-of-plane rotations are about the minimum principal inertia axis; the secondary's angular momentum is still dominated by its rotation about the maximum principal inertia axis. Therefore the system is not far from equilibrium. Furthermore, the excess angular momentum goes into the out-of-plane rotation instead of in-plane libration. Therefore, we expect the post-impact libration in this scenario to be similar to the planar case, i.e. relatively unaffected by the initial out-of-plane rotation. So we do not expect initial out-of-plane rotations to have a significant effect on the dynamics, at least for small values of $\beta$. A more important consequence of initial out-of-plane rotation is it becomes easier for the impact to further excite this rotation mode, known as the barrel instability \citep{cuk2021barrel}, and the secondary may start tumbling at lower values of $\beta$ than expected otherwise.
 
\subsection{Constituent Modes of Variations}\label{sec:modes} 
\begin{figure}[ht]
   \centering
   \includegraphics[width = 3in]{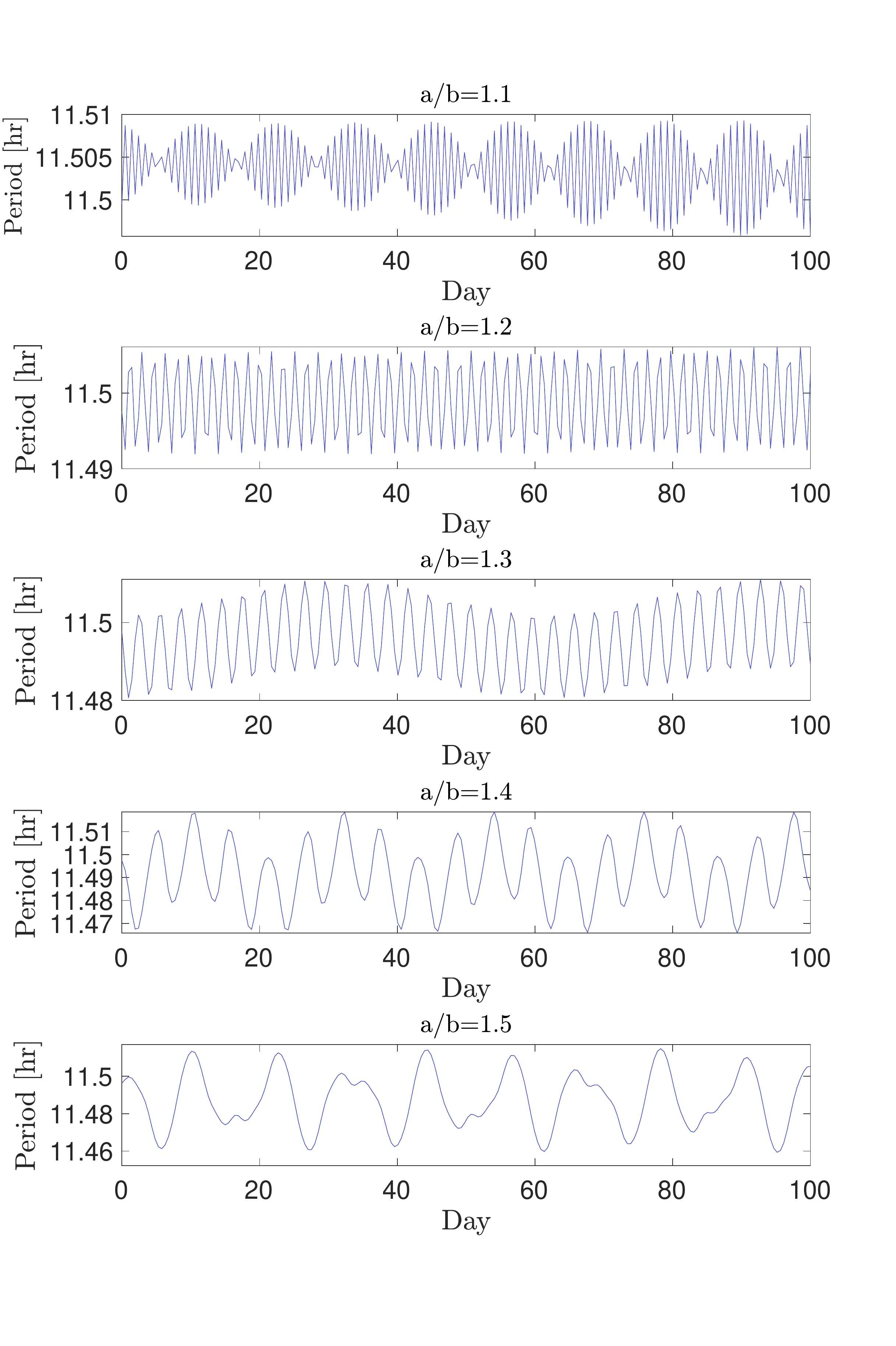} 
   \caption{The instantaneous orbit period for different values of $a/b$, all for $\beta=3$. The orbit period variation behavior strongly depends on $a/b$.}
   \label{period_all_ab}
\end{figure}

As shown in Section \ref{sec:analytic}, the orbit period variations are driven by two modes: a short-period and long-period. However, the constituent modes are affected by both $a/b$ and $\beta$. An example of this is shown in Figure \ref{period_all_ab}, where the instantaneous orbit period is plotted over time for $a/b$ varying between 1.1 and 1.5, all for $\beta=3$. Each axis ratio shows a unique behavior in the orbit period, highlighting the strong dependence on the secondary's shape.

To compare the behavior of the orbit period variations, we compute the discrete Fourier transform of each time history and compare their power spectral densities to obtain estimates of the period and amplitude of the constituent modes. The Fourier transform struggles to obtain accurate values of the frequencies owing to the discretization of the orbit period. Since only one data point is obtained per orbit period, the sampling frequency is necessarily course. To counteract this we use a long time domain of 500 days. However, the results are only estimates of the true behavior due to this shortcoming.

Table \ref{ab_modes} shows the results of the Fourier transform. The period and amplitude estimates of the short-period and long-period modes are reported for values of $a/b$. Note that the transform did not find a long-period mode for $a/b=1.2$. Additionally, a third mode for $a/b=1.1$ causes beating, with a period of approximately 11 days. Recall the resonance at $a/b=1.1$, which may explain this unique behavior in the orbit period variations. In general, Table \ref{ab_modes} suggests increasing $a/b$ will increase the short period. The short-period amplitude also increases with $a/b$ until after $a/b=1.4$, where it suddenly decreases. Note the inclusion of $a/b=1.25$, which is approximately the longest mode with non-negligible amplitude. The DART mission must measure the post-impact period to an accuracy of 7.3 seconds \citep{cheng2018aida}, so observations must span for the length of any long-period mode with amplitude larger than $2\times7.3=14.6$ seconds. For $\beta=3$, any long-period mode for $a/b=1.24$ or less has an amplitude below this threshold. The long-period mode amplitude has its minimum at $a/b=1.2$ where it is non-existent, then increases in both directions as $a/b$ changes. The period behaves oppositely, where it can be thought of as infinitely long at $a/b=1.2$ and decreases as $a/b$ changes in either direction. In this way, any $a/b$ close to 1.2 will have a very small long-period mode amplitude with a very long period.

\begin{table}[ht]
\begin{center}
 \caption{\label{ab_modes} The period and amplitude of the short-period and long-period constituent modes of the orbit period variations for values of $a/b$, all using $\beta=3$. These estimates are obtained through a discrete Fourier transform of the instantaneous orbit period time histories.}
\begin{tabular}{c|cccc } 

 & \multicolumn{2}{c}{Short-period mode} & \multicolumn{2}{c}{Long-period mode} \\
$a/b$ & period & amplitude & period & amplitude\\
\hline
1.1 & 1 day & 48.4 s & 250 day & 8.6 s \\
1.2 & 1.7 day & 63.0 s & N/A & N/A \\
1.25 & 2.2 day & 63.4 s & 125 day & 19.2 s \\
1.3 & 3.0 day & 83.9 s & 62.0 day & 31.1 s \\
1.4 & 5.4 day & 127.3 s & 21.8 day & 70.7 s \\
1.5 & 6.8 day & 68.0 s & 11.4 day & 136 s \\

 \hline
 \end{tabular}
 \end{center}
\end{table}

Next, we repeat the same process but holding $a/b=1.3$ constant and varying $\beta$ from 1 to 5. The estimates for the amplitudes and periods of the modes are shown in table \ref{beta_modes}. As expected, increasing $\beta$ increases the amplitudes of both the short- and long-period modes. The results also suggest that increasing $\beta$ decreases the short period, while the long period increases.

\begin{table}[ht]
\begin{center}
 \caption{\label{beta_modes} The period and amplitude of the short-period and long-period constituent modes of the orbit period variations for values of $\beta$, all using $a/b=1.3$. These estimates are obtained through a discrete Fourier transform of the instantaneous orbit period time histories.}
\begin{tabular}{c|cccc } 

 & \multicolumn{2}{c}{Short-period mode} & \multicolumn{2}{c}{Long-period mode} \\
$\beta$ & period & amplitude & period & amplitude\\
\hline
1 & 3.4 day & 32.4 s & 54 day & 14.3 s \\
2 & 3.2 day & 56.7 s & 58 day & 26.9 s \\
3 & 3.0 day & 84.7 s & 62 day & 30.8 s \\
4 & 2.8 day & 103.7 s & 70 day & 37.0 s \\
5 & 2.6 day & 130.9 s & 88 day & 40.1 s \\

 \hline
 \end{tabular}
 \end{center}
\end{table}

The short- and long-period modes appear to be driven by the apsidal precession of the eccentricity vector, which is measured by the longitude of periapsis. On average, the longitude of periapsis precesses with a frequency equal to the long-period mode in the orbit period variations. Within that precession, the eccentricity vector also sees short-period oscillations, with a period equal to the short-period mode in the orbit period variations. Interestingly, the eccentricity \textit{magnitude} oscillates at the same frequency as the separation and libration, which is not the same as the short-period mode in the orbit period variations. Instead, the orbit period variations are driven by the \textit{orientation} of the eccentricity vector as the elliptical orbit precesses.

While this provides a physical interpretation of the frequencies, it is desirable to obtain an analytic approximation of the frequencies that drive the orbit period variations. To accomplish this, we compute an equilibrium solution at the new value of free angular momentum, $H_{free}^*$, giving us an equilibrium separation $r^*$ and orbital frequency $n^*$. We then treat the post-impact system as a perturbed solution around this new equilibrium and compute the corresponding linearized solution. The variations in $r$ and $\lambda$ can be expressed as:

\begin{equation}
    \delta r(t) = A_r\cos\omega_1t+B_r\cos\omega_2t
\end{equation}
\begin{equation}
    \delta\lambda(t) = A_\lambda\cos\omega_1t+B_\lambda\cos\omega_2t ,
\end{equation}
where the coefficients $A_i$ and $B_i$ are constants that depend on the system parameters. However, we are mostly concerned with the frequencies, so we focus on $\omega_1$ and $\omega_2$, which are the linearized frequencies driving oscillations in $r$ and $\lambda$. These frequencies can be found as the solution of the biquadratic equation:

\begin{equation}
    \xi^4+\alpha\xi^2+\gamma=0,
\end{equation}
with

\begin{equation}
    \alpha = \frac{3\mathcal{G}(m_A+m_B)(m_B(I_{s}-I_{1z})+m_A(I_{2y}-I_{2z}))}{2m_Am_Br^{*2}}+\frac{\mathcal{G}(m_A(-3I_{2x}+3I_{2y}+I_{2z})+m_BI_{2z})}{I_{2z}r^{*3}}
    \label{alpha_coef}
\end{equation}
and
\begin{multline}
    \gamma = \frac{-3\mathcal{G}^2(I_{2x}-I_{2y})(m_A+m_B)}{2I_{2z}m_Am_B^2r^{*10}}\bigg(-3m_Am_Br^{*2}\big(m_B(I_{1z}+2I_{2z}-I_s)+m_A(-2I_{2x}+I_{2y}+3I_{2z})\big) \\-15I_{2z}(m_A+m_B)\big(m_B(I_{1z}-I_{s})+m_A(-2I_{2x}+I_{2y}+I_{2z})\big)+2m_A^2m_B^2r^{*4}\bigg).
    \label{gamma_coef}
\end{multline}
The derivation of these coefficients is outlined in the \nameref{sec:appendix}. We now turn to the orbit period, and approximate the orbital displacement as
\begin{equation}
    \theta(t)=n^*t+A_1\cos\omega_1t+A_2\cos\omega_2t.
\end{equation}
Expressing time as $t=pT_0+\delta T$ where $T_0=2\pi(n^*)^{-1}$ and $\delta T$ is the variable deviation from this value, we obtain an approximate expression for $\theta(t)$ ignoring second-order terms:

\begin{equation}
    \theta(t)=2p\pi+A_1\cos\frac{2p\pi\omega_1}{n^*}+A_2\cos\frac{2p\pi\omega_2}{n^*}+\delta Tn^*,
\end{equation}
where $p$ is an integer number of orbits. Setting the approximate $\theta(t)=2p\pi$ allows us to compute the correction $\delta T$ after $p$ orbits:
\begin{equation}
    \delta T(p) \approx \frac{A_1\cos\frac{2p\pi\omega_1}{n^*}+A_2\cos\frac{2p\pi\omega_2}{n^*}}{-n^*}.
\end{equation}

A Fourier analysis of this approximation shows that the dominant frequencies driving the orbit period variations are approximately $(n^*-\omega_1)$ and $(n^*-\omega_2)$. Owing to the linearization, this approximation is only accurate at small values of $\beta$, and becomes less accurate as $\beta$ increases. However, it provides a good first approximation of the frequencies that drive the orbit period variations. For example, for the axis ratio $a/b=1.3$, at $\beta=1$ this method calculates the short period equal to 3.4 days and the long period equal to 50 days. This is more accurate than the $\beta=3$ estimate, which calculates the short period equal to 3.3 days and the long period equal to 48 days. Note the short-period mode remains relatively accurate at larger $\beta$, but the long-period mode has the incorrect dependence on $\beta$. Nevertheless, this approach provides an analytic approximation of the frequencies of orbit period variations. This also reveals that the long-period mode, which is driven by the precession of the elliptical orbit, is strongly affected by nonlinear effects.

\section{High-Fidelity Simulations} \label{sec:high-fidelity}
Up to now we have relied on the simplified model of \cite{mcmahon2013dynamic} to simulate the system dynamics, with only a few comparisons to high-fidelity simulations to validate our results. We now turn to \textsc{gubas} to run high-fidelity simulations, using full three-dimensional dynamics and polyhedron shape models for Didymos and Dimorphos. While a shape model exists for Didymos, there is none for Dimorphos. In this section we aim to avoid the ellipsoidal assumption on Dimorphos, since in reality it is unlikely to be symmetric and we wish to test the effect of an asymmetric secondary. We therefore use the shape model of Squannit, the secondary in the Moshup binary asteroid system, scaled to the expected size of Dimorphos. We adjust the shape model such that its volume is equivalent to the expected size of Dimorphos and its axis ratios are approximately $a/b=1.3$ and $b/c=1.2$.

\begin{figure}[ht]
   \centering
   \includegraphics[width = 3in]{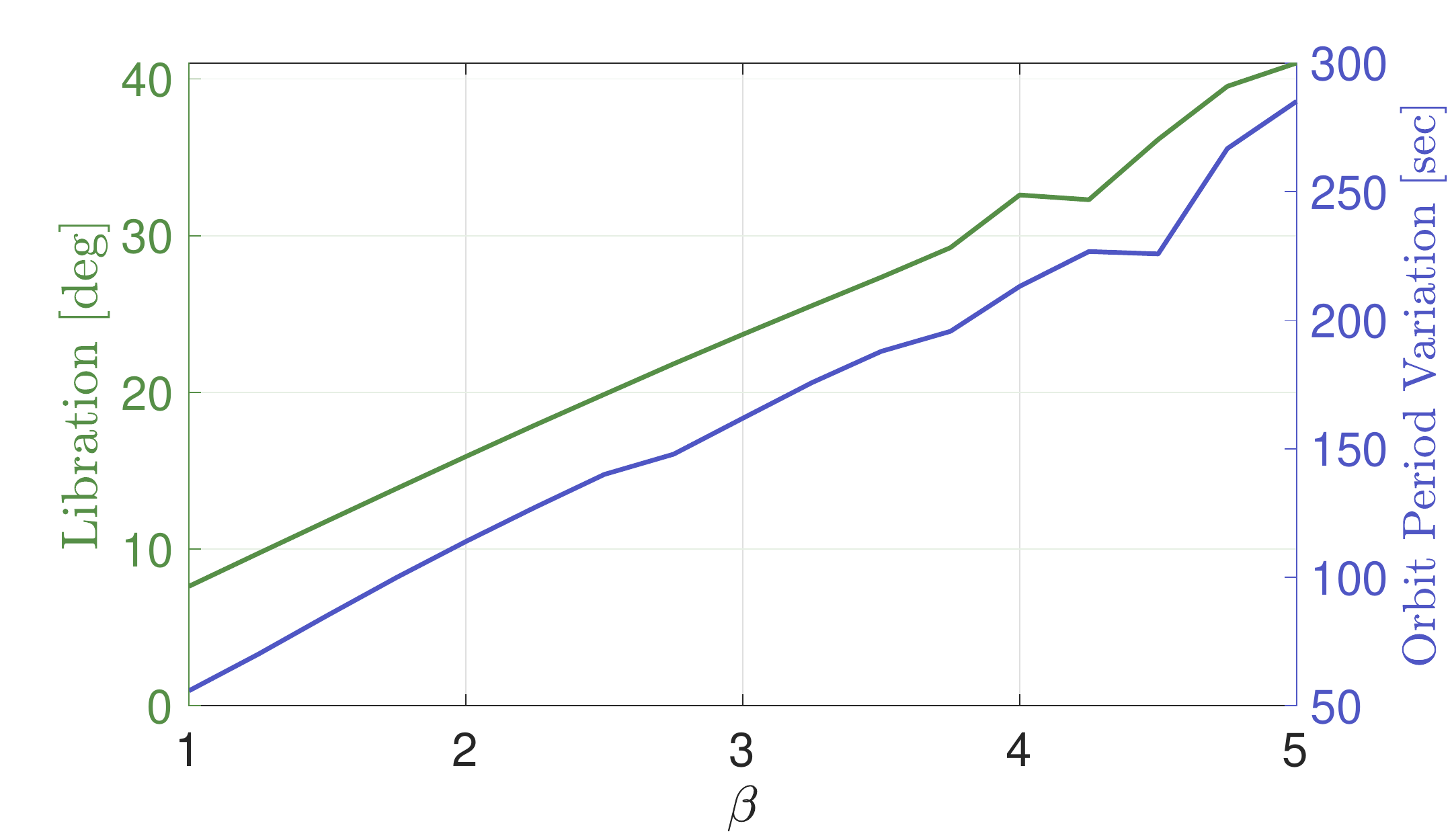} 
   \caption{The libration and orbit period variation of the full 3D \textsc{gubas} simulations using a stand-in polyhedron model for Dimorphos.}
   \label{beta_lib_period_GUBAS}
\end{figure}

We use a more realistic three-dimensional impact geometry, where the $\Delta v$ from the DART impact is applied at a vector approximately $10^\circ$ `up' out of plane and $10^\circ$ `out' in the radial direction. We again test values of $\beta$ ranging from 1 to 5. Figure \ref{beta_lib_period_GUBAS} shows the resulting libration and orbit period variation magnitudes for each simulation. Note that while the libration magnitudes are slightly larger than than simplified cases, the orbit period sees much larger variations. This demonstrates how the out-of-plane dynamics are important, as well as the coupled dynamics between more complicated shape models. These results also highlight the importance of obtaining an accurate estimate of Dimorphos's shape and mass to model its dynamics.

\begin{figure}[ht]
   \centering
   \includegraphics[width = 3in]{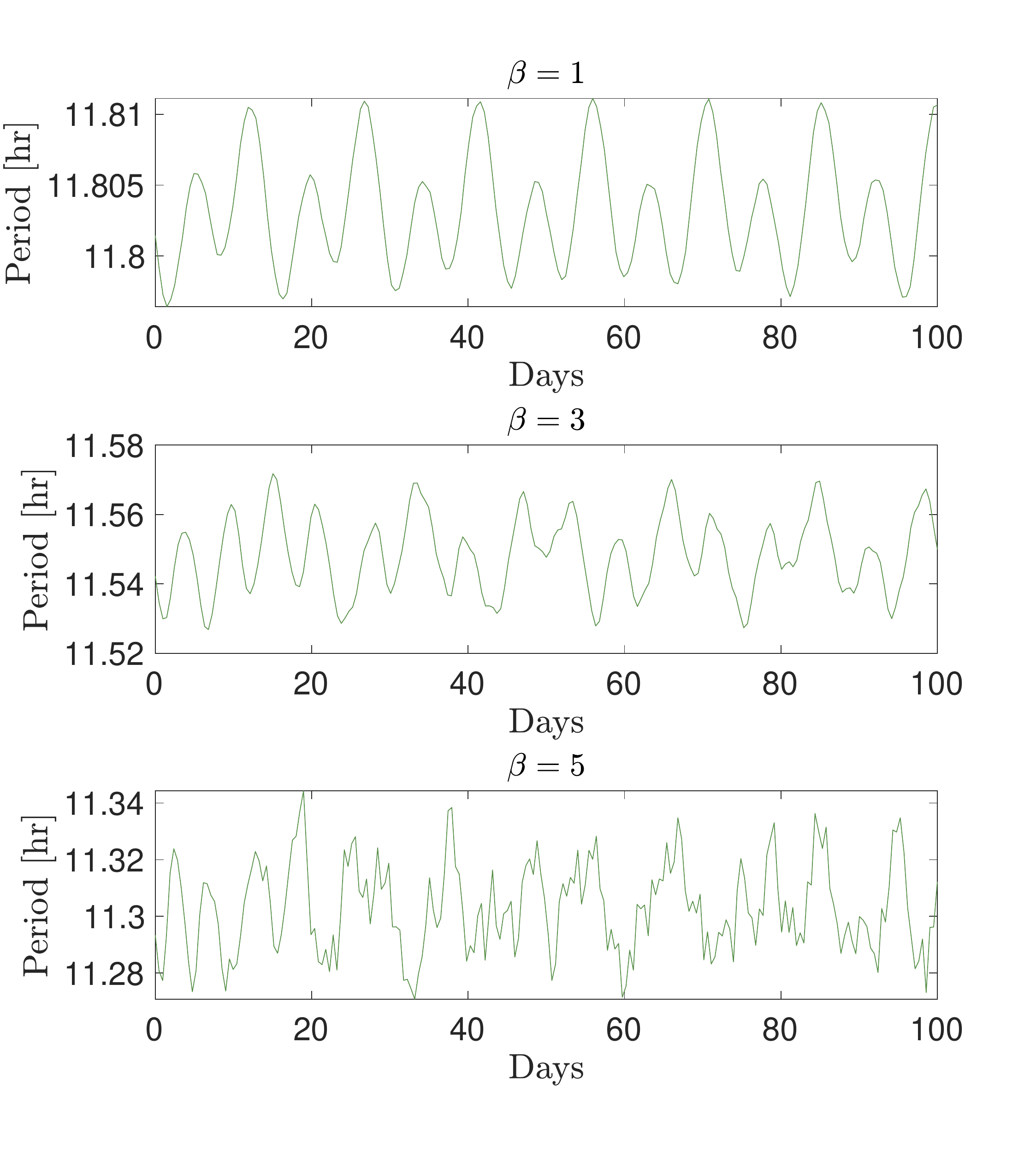} 
   \caption{The time history of the orbit period for $\beta=1,3,$ and 5 from the full 3D \textsc{gubas} simulations. At $\beta=5$ the two-mode structure breaks down and the orbit period variations become irregular. This corresponds to out-of-plane rotation in the barrel instability.}
   \label{period_all_ab_GUBAS}
\end{figure}

Time histories of the orbit period for the $\beta = 1,3,$ and 5 simulations are plotted in Figure \ref{period_all_ab_GUBAS}. Here we see the two-mode structure appear in the $\beta=1$ and 3 cases, but not for $\beta=5$. At $\beta=5$, the impact transfers enough momentum to excite the barrel instability. This more complex rotation destroys the two-mode structure we saw in the planar cases. Again this illustrates the importance of considering the out-of-plane dynamics. However, this also shows that the simplified planar model is a fairly good prediction of the dynamics at low values of $\beta$.

\section{Observation Implications} \label{sec:observation}
Due to the coupled relationship between the libration amplitude and the fluctuations in the orbit period, if one of these characteristics can be measured it can offer insight into the other, provided an estimate of the secondary's shape and mass is available. One possible application of this is constraining the libration state of the secondary through observations of the orbit period. Since the orbit period is easier to measure from the ground than the libration state, there is great potential in this approach. However, capturing the orbit period variations to sufficient accuracy may be difficult.

More relevant to the DART mission, however, is measuring the mean post-impact orbit period. To explore this process we turn to the nominal case, $a/b=1.3$, $b/c=1.2$, and $\beta=3$. This is the case plotted in Figures \ref{analytic_results} and \ref{analytic_results_zoom}. As discussed in depth, the orbit period fluctuations are driven by two frequencies: a short-period and a long-period. The short-period frequency has a period of about 3 days, while the long-period frequency has a period on the order of a few months. To capture the full orbit period variation, both these modes must be accounted for, as both of their amplitudes contribute to the overall variations in the orbit period.

\newpage
\subsection{Photometric Observations}
To examine the modulation of the mean anomaly for photometric observation purposes, we model the instantaneous mean motion as a sum of two modes:

\begin{equation}
n(t) = \bar{n} + A_1\cos(\omega_1t-\phi_1) + A_2\cos(\omega_2t-\phi_2).
\end{equation}

\begin{figure}[h]
   \centering
   \includegraphics[width = 3in]{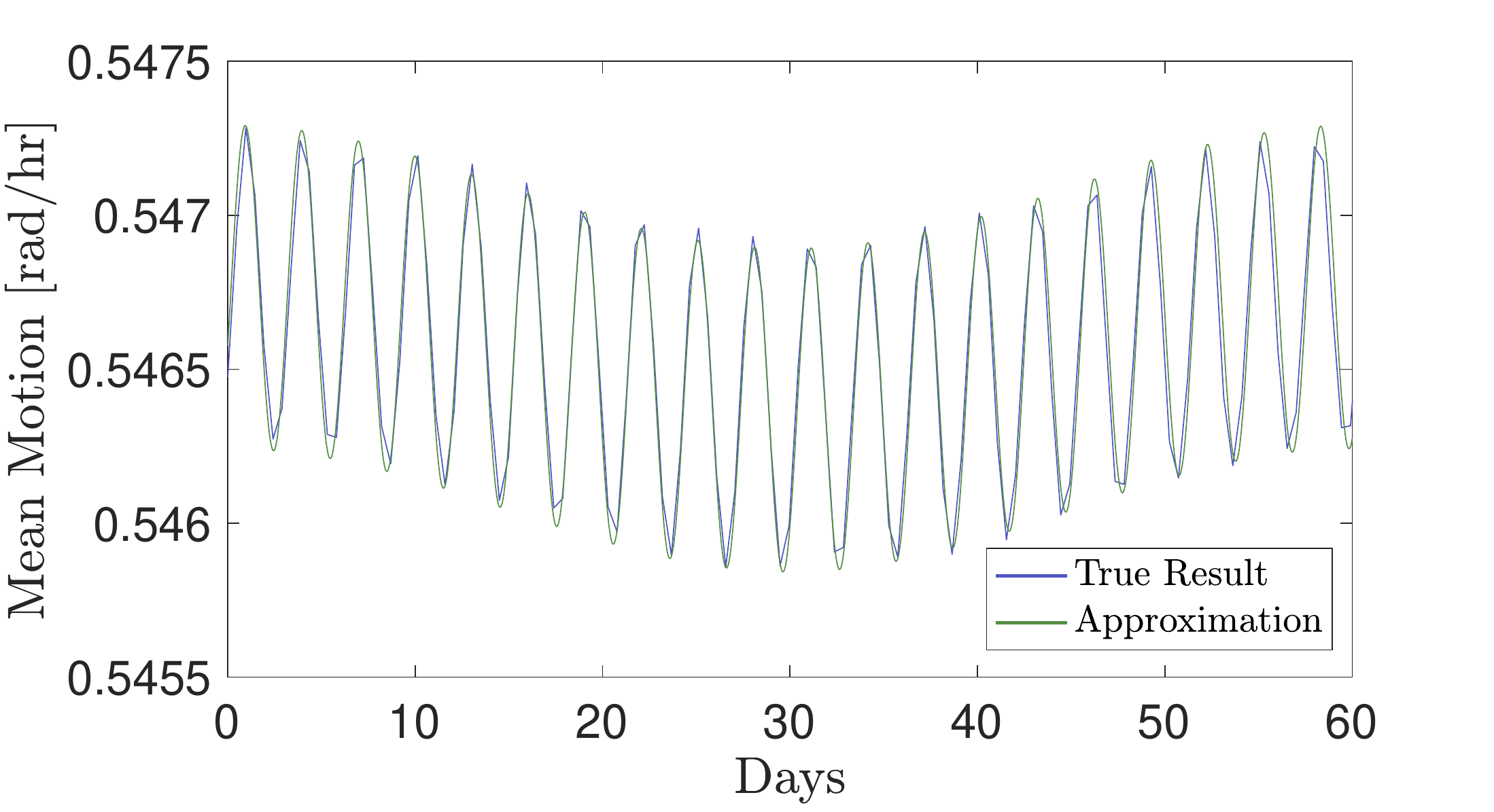} 
   \caption{The instantaneous mean motion with the analytic approximation.}
   \label{mean_motion}
\end{figure}

The average mean motion is about $\bar{n}=1.518\times10^{-4}$ rad/s. The short-period mode has amplitude about $A_1 = 1.458\times10^{-7}$ rad/s and frequency about $\omega_1=2.408\times10^{-5}$ rad/s. The long-period mode has amplitude about $A_2 = 5.556\times10^{-8}$ rad/s and frequency about $\omega_2=1.212\times10^{-6}$ rad/s. The approximation is plotted with the actual mean motion variation in Figure \ref{mean_motion}, showing a close match.

This gives a mean anomaly modulation amplitude of the short-period equal to $A_1/\omega_1=0.35$ deg and long-period modulation amplitude equal to $A_2/\omega_2=2.63$ deg. Since the short-period modulation of the mean anomaly is smaller than uncertainties in the event timings from other sources, mainly noise in the photometry measurements and irregularities in the asteroid shapes, it is negligible and can be considered as simply an additional source of noise. However, the long-period modulation is comparable to the event timing uncertainties from other sources, and so must be considered. It is therefore necessary to spread observations over a few months to capture the long-period variation and ensure an accurate estimate of the mean orbit period.

\subsection{Radar Observations}
Radar observations of Didymos are planned at Goldstone and Green Bank starting in late September 2022 and lasting for up to three weeks. If the observations are bistatic, as planned, then the signal-to-noise ratios in 2022 will be about half as strong as they were during monostatic Goldstone observations in 2003, and as a result, the finest range resolution will probably be 75 meters per pixel. This will show the positions of the primary and secondary at different epochs and will be sufficient to spatially resolve the primary but will place only a few pixels on the secondary due to its relatively small size.

Due to the weak radar SNRs expected in 2022, it is very unlikely that the delay-Doppler images will have sufficient spatial resolution to reveal subtle variations in the orientation of the secondary at different epochs other than due to its orbital position. Librations will probably not be visible in the radar data.

It seems likely the DART impact will cause variations superimposed on the impulsive change to the orbit period. The average change could become visible in the positions of Dimorphos within one week and possibly even more rapidly. The magnitude of the short-period semimajor axis variations are expected to be roughly 60 meters, which is comparable to the finest range resolution provided by the radar observations. As such, it seems unlikely that variations with these periods will be evident in the radar imaging data, and at best, perhaps a periodic variation of 3 days \textit{might} be visible in the residuals to orbital fits. The radar observations might span about three weeks, so it seems even less likely that the long-period mode will be visible. However, the uncertainties are considerable so the data will be checked thoroughly just in case.

The variations in the orbit period are challenging in the context of the DART mission. One of the main requirements of the DART mission is to measure the change in orbit period to a precision of 7.3 seconds \citep{cheng2018aida}. If the orbit period is changing over time, it may be more difficult to meet this level of precision than originally thought due to the additional noise on the mean orbit period generated by the variations. However, measuring the mean orbit period to within the required precision is possible despite the variations. Furthermore, provided the libration is not damped over the 4 years it will take for the arrival of Hera at Didymos, Hera may be able to directly measure the orbit period variations.

\section{Discussion and Conclusion} \label{sec:discussion}
The libration of the secondary is not independent of the mutual orbit dynamics in a binary asteroid system. Due to the conservation of angular momentum, fluctuations in the secondary's spin state cause a response in the orbital angular momentum. As a result, the orbital period is non-constant in a librating binary asteroid system. If a system is in equilibrium with no libration, the orbit period is constant, but any deviations from that equilibrium will result in variations in the orbit period. There is a real-world example of a changing orbit period in binary asteroids. Variations appear in 1991 VH, where there is likely an interaction between the secondary's spin state and the orbit dynamics \citep{Pravec2021Photometric, meyer2021modeling}.

We see the orbit period variations are driven by two modes, both of which are have a significant amplitude. The variations are strongly dependent on $\beta$ and the secondary axis ratio $a/b$, with certain values of $a/b$ causing resonances in the dynamics. These modes are driven by the precession of the elliptical orbit and the orientation of the eccentricity vector.

The orbit period variations have significant implications for the observation of Didymos following the impact. It is unlikely the short-period mode in the variations will be observable from the ground, but the long-period mode may need to be accounted for in observations. This offers an additional challenge for the DART mission, where measuring the post-impact orbit period may be more difficult due to the variations. Note this challenge will benefit from the complementarity of the Hera mission that will measure the post-impact state in great detail in the rendezvous with the target 4 years later. Despite the difficulties provided by the orbit period variations, the mean orbit period change will still be measurable to the required accuracy.

If the secondary's shape is also known along with the initial system configuration and orientation at impact, the post-impact dynamics can provide an additional constraint on the estimation of $\beta$ instead of relying solely on the mean period change. Thus, while orbit period variations are certainly a challenge in measuring the period, they can also provide an opportunity to gain more information on the impact. And since the orbit period variations are closely tied to the secondary's libration, they can also help constrain the libration state given proper knowledge of the secondary's shape and mass. This provides additional motivation for LICIACube to accurately image the system just before the DART impact.

\cite{agrusa2021excited} show that equilibrium configurations of Dimorphos are extremely sensitive to shape changes. If Dimorphos is a rubble pile asteroid, its shape might be subject to small changes and medium-term adjustments, especially after the DART impact. In this case, the characteristic time of the shape change will be comparable to the orbital and spin period of Dimorphos. This is confirmed by preliminary results of rubble-pile simulations run using DEM codes \textsc{pkdgrav}~\citep{Richardson2000} and \textsc{grains}~\citep{Ferrari2017}, which show that even small shape change can lead to strong oscillations in Dimorphos's libration motion. Compared to the rigid body case, the slow but steady deformation of the body facilitates the motion between stable and unstable regions identified by \cite{agrusa2021excited}. As discussed, in a binary system libration motion is closely connected to the orbital motion of the secondary, meaning that stronger libration oscillations reflects in larger fluctuations in the orbit period as well. A more detailed discussion of the rubble-pile case is provided in a dedicated paper, currently in preparation.

Most of this work assumed a planar impact through the secondary's center of mass. In reality, the impact will not be planar, as DART will impact Dimorphos at an angle of about $10^{\circ}$ from the orbit plane \citep{cheng2018aida}. Furthermore, it is likely that the actual impact will not be coincident with the center of mass in reality. Due to these simplifications, the analysis in this study likely does not capture all of the relevant dynamics, and the actual results are likely to be more complicated than predicted here. While Section \ref{sec:high-fidelity} relaxed these assumptions for a more accurate analysis, it is by no means comprehensive. These simplifications were made to demonstrate the coupling between libration and orbit period variations, but the full non-planar dynamics should be considered in future analysis of the impact, along with the torque imparted by the spacecraft on Dimorphos. An off-center impact will likely induce a higher libration magnitude, resulting in larger variations in the orbit period. The in-situ dynamical and physical characterization of the target by Hera will allow us to ultimately remove the possible ambiguities in the impact.

\section*{Acknowledgements}
We would like to thank the entire DART investigation team for helpful discussion throughout this study's development and preparation. This study was supported in part by the DART mission, NASA Contract \#NNN06AA01C to JHU/APL. The work of P.P. was supported by the Grant Agency of the Czech Republic, Grant 20-04431S. P.M. acknowledges support from CNES and ESA. I.G., M.G., K.T., and P.M. acknowledge funding from the European Union’s Horizon 2020 research and innovation program under grant agreement No. 870377 (project NEO-MAPP).

\section*{Appendix} \label{sec:appendix}
\renewcommand{\theequation}{A\arabic{equation}}
\setcounter{equation}{0}
To derive the analytic approximations of the linear frequencies we begin with the Hamiltonian of the planar model:
\begin{equation}
    \mathcal{H} = \frac{1}{2}\left( \frac{p_\lambda}{I_{Bz}}+\frac{(p_\theta-p_\lambda)^2}{r^2\mu}+\frac{p_r^2}{\mu} \right) + V(r,\lambda)
\end{equation}
where the unnormalized potential is
\begin{equation}
    V(r,\lambda)=-\frac{\mathcal{G}m_Am_B}{r}-\frac{\mathcal{G}m_B}{2r^3}(2I_s+I_{Az})-\frac{\mathcal{G}m_A}{2r^3}(I_{Bx}+I_{By}+I_{Bz})+\frac{3\mathcal{G}m_B}{2r^3}I_s+\frac{3\mathcal{G}m_A}{4r^3}\bigg(I_{Bx}+I_{By}-(I_{By}-I_{Bx})\cos2\lambda\bigg).
\end{equation}
This leads to the equations of motion:
\begin{equation}
\label{flow}
    \dot{r}=\frac{\partial \mathcal{H}}{\partial p_r},
\quad
    \dot{p_r}=-\frac{\partial \mathcal{H}}{\partial r},
\quad
    \dot{\lambda}=\frac{\partial \mathcal{H}}{\partial p_\lambda},
\quad
    \dot{p_\lambda}=-\frac{\partial \mathcal{H}}{\partial \lambda},
\end{equation}
where
\begin{equation}
    p_r=\frac{\partial \mathcal{L}}{\partial \dot{r}},
\quad
    p_\lambda=\frac{\partial \mathcal{L}}{\partial \dot{\lambda}},
\end{equation}
and $\mathcal{L}$ is the Lagrangian of the planar system. From \cite{mcmahon2013dynamic}, the binary is in an equilibrium configuration when
\begin{equation}
    n^*=\sqrt{\frac{\mu}{r^{*3}}\bigg( 1+\frac{3}{2r^{*2}}\bigg( \frac{I_{Az}-I_s}{m_A} + \frac{I_{By}+I_{Bz}-2I_{Bx}}{m_B} \bigg) \bigg)},
\end{equation}
which also means
\begin{equation}
    \lambda=\dot{\lambda}=0.
\end{equation}
Following the DART impact, we calculate the new angular momentum and the new equilibrium $r^*$ and $n^*$ that correspond to this angular momentum. We treat the initial condition (at impact) as a perturbed solution around the new equilibrium and compute the corresponding linearized solution. We introduce a deviation vector
\begin{equation}
    \vec{w}=[\delta r, \delta \lambda, \delta p_r, \delta p_\lambda]
\end{equation}
and the linearized system
\begin{equation}
    \dot{\vec{w}}=\mathcal{J}\vec{w},
    \label{linearized}
\end{equation}
where $\mathcal{J}$ is the Jacobian of the flow Equations \ref{flow}. Substituting the new equilibrium condition $(r^*,n^*)$ into Equation \ref{linearized}, we obtain the characteristic polynomial
\begin{equation}
    \xi^4+\alpha\xi^2+\gamma=0,
\end{equation}
here the coefficients $\alpha$ and $\gamma$ are as defined in Equations \ref{alpha_coef} and \ref{gamma_coef}. The linearized frequencies $\omega_1$ and $\omega_2$ are then given as the solutions of the characteristic polynomial.

\bibliography{References}{}
\bibliographystyle{aasjournal}

\end{document}